\documentclass[twocolumn]{aastex631}

\received{April 15, 2024}
\accepted{\today}
\submitjournal{ApJL}
\shorttitle{Perilous parabolic passages of polytropes}
\shortauthors{Price et al.}

\begin{document}

\title{Eddington envelopes: The fate of stars on parabolic orbits tidally disrupted by supermassive black holes}

\correspondingauthor{Daniel J. Price}
\author[0000-0002-4716-4235]{Daniel J. Price}
\affiliation{School of Physics and Astronomy, Monash University, Clayton, Vic. 3800, Australia}
\email{daniel.price@monash.edu}

\author[0000-0002-4856-1750]{David Liptai}
\affiliation{School of Physics and Astronomy, Monash University, Clayton, Vic. 3800, Australia}

\author[0000-0002-6134-8946]{Ilya Mandel}
\affiliation{School of Physics and Astronomy, Monash University, Clayton, Vic. 3800, Australia}
\affiliation{The ARC Centre of Excellence for Gravitational Wave Discovery -- OzGrav}
\affiliation{Birmingham Institute for Gravitational Wave Astronomy and School of Physics and Astronomy, University of Birmingham,  B15 2TT, Birmingham, UK}

\author[0000-0001-8180-8842]{Joanna Shepherd}
\affiliation{School of Physics and Astronomy, Monash University, Clayton, Vic. 3800, Australia}

\author[0000-0002-2357-7692]{Giuseppe Lodato}
\affiliation{Dipartimento di Fisica, Universit\`a Degli Studi di Milano, Via Celoria 16, Milano 20133, Italy}

\author[0000-0002-6987-1299]{Yuri Levin}
\affiliation{Center for Theoretical Physics, Department of Physics, Columbia University, New York, NY 10027, USA}
\affiliation{Center for Computational Astrophysics, Flatiron Institute, New York, NY 10010, USA}
\affiliation{School of Physics and Astronomy, Monash University, Clayton, Vic. 3800, Australia}

\begin{abstract}
Stars falling too close to massive black holes in the centres of galaxies can be torn apart by the strong tidal forces. Simulating the subsequent feeding of the black hole with disrupted material has proved challenging because of the range of timescales involved.
Here we report a set of simulations that capture the relativistic disruption of the star, followed by one year of evolution of the returning debris stream. These reveal the formation of an expanding asymmetric bubble of material extending to hundreds of astronomical units --- an outflowing Eddington envelope with an optically thick inner region.
Such outflows have been hypothesised as the reprocessing layer needed to explain optical/UV emission in tidal disruption events, but never produced self-consistently in a simulation.
Our model broadly matches the observed light curves with low temperatures, faint luminosities, and line widths of 10,000--20,000 km/s.
\end{abstract}


\section{Introduction} \label{sec:intro}
In the classical picture of tidal disruption events (TDEs), the debris from the tidal disruption of a star on a parabolic orbit by a supermassive black hole (SMBH) rapidly circularises to form an accretion disc via relativistic apsidal precession \citep{rees88}.
The predicted mass return rate of debris \citep{phinney89} is $\propto t^{-5/3}$ and the light curve is assumed to be powered by accretion and to follow the same decay.

This picture alone does not predict several properties of observed TDEs, mainly related to their puzzling optical emission \citep{van-velzenfarrargezari11,van-velzen18,van-velzenetal21}.
These properties include:
i) low peak bolometric luminosities \citep{chornockbergergezari14} of $\sim 10^{44}$ erg/s, $\sim\,$1 per cent of the value expected from radiatively efficient accretion \citep{svirskipirankrolik17};
ii) low temperatures, more consistent with the photosphere of a B-type star than with that of an accretion disc at a few tens of gravitational radii ($R_g\equiv GM_\mathrm{BH}/c^2$) \citep{gezarichornockrest12,miller15}, and consequently large emission radii, $\sim\,$$10$--$100$ au for a $10^6 M_\odot$ black hole  \citep{guillochonmanukianramirez-ruiz14,metzgerstone16}; and
iii) spectral line widths implying gas velocities of $\sim\,$$10^4$ km/s, much lower than expected from an accretion disc \citep{arcavigal-yamsullivan14,leloudasdaiarcavi19,Nicholl19}. 

As a consequence, numerous authors have proposed alternative mechanisms for powering the TDE lightcurve, via either shocks from tidal stream collisions during disc formation  \citep{lodato12,piransvirskikrolik15,svirskipirankrolik17,ryuetal23,huangetal23}, or the reprocessing of photons through large scale optically thick layers, referred to as Eddington envelopes \citep{loebulmer97}, super-Eddington outflows \citep{strubbequataert09}, quasi-static or cooling TDE envelopes \citep{rothkasenguillochon16,coughlinbegelman14,metzger22} or mass-loaded outflows  \citep{jiangguillochonloeb16,metzgerstone16}. Recent spectro-polarimetric observations  suggest reprocessing in an outflowing, quasi-spherical envelope \citep{patraetal22}.

The wider problem is that few calculations exist that follow the debris from disruption to fallback for a parabolic orbit with the correct mass ratio. The challenge is to evolve a main-sequence star on a parabolic orbit around a SMBH from disruption and to follow the subsequent accretion of material \citep{metzgerstone16}.
The dynamic range involved when a $1M_\odot$ star on a parabolic orbit is tidally disrupted by a $10^6 M_\odot$ SMBH is greater than four orders of magnitude: the tidal disruption radius is 50 times the gravitational radius, where general relativistic effects are important, while the apoapsis of even the most bound material is another factor of 200 further away.
This challenge led previous studies to consider a variety of simplifications \citep{stonekesdencheng19}:
i) reducing the mass ratio between the star and the black hole by considering intermediate mass black holes \citep{ramirez-ruizrosswog09,guillochonmanukianramirez-ruiz14};
ii) using a Newtonian gravitational potential \citep{evanskochanek89,rosswogramirez-ruizhix08,lodatokingpringle09,guillochonramirez-ruizrosswog09,golightlycoughlinnixon19}, pseudo-Newtonian \citep{hayasakistoneloeb13,bonnerotrossilodato16} or post-Newtonian approximations \citep{ayalliviopiran00,hayasakistoneloeb16};
iii) simulating only the first passage of the star  \citep{evanskochanek89,lagunamillerzurek93,khokhlovnovikovpethick93,frolovkhokhlovnovikov94,dienerfrolovkhokhlov97,kobayashilagunaphinney04,guillochonramirez-ruizrosswog09,guillochonramirez-ruiz13,tejedagaftonrosswog17,gaftonrosswog19,goicovicspringelohlmann19};
and iv) assuming stars initially on bound, highly eccentric orbits instead of parabolic orbits  \citep{sadowskitejedagafton16,hayasakistoneloeb13,hayasakistoneloeb16,bonnerotrossilodato16,liptaipricemandel19,huetal24}.

 These studies have, nevertheless, provided useful insights into the details of the tidal disruption process.
In particular, it has been shown that the distribution of orbital energies of the debris following the initial disruption is roughly consistent with $dM/dE = $const, consistent with the analytic prediction of a $\propto t^{-5/3}$ mass fallback rate, although the details can depend on many factors such as stellar spin, stellar composition, penetration factor and black hole spin  \citep{lodatokingpringle09,kesden12,guillochonramirez-ruiz13,golightlycoughlinnixon19,sacchilodato19}.
The importance of general relativistic effects in circularising debris has also been demonstrated.
The self-intersection of the debris stream, which efficiently dissipates large amounts of orbital energy, is made possible by relativistic apsidal precession \citep[][]{hayasakistoneloeb16,bonnerotrossilodato16,liptaipricemandel19,calderonetal24}.
But until recently debris circularisation has only been shown for stars on bound orbits, with correspondingly small apoapsis distances and often deep penetration factors (we define the penetration factor as $\beta\equiv R_\mathrm{t}/R_\mathrm{p}$,
where $R_\mathrm{t}=R_*(M_\mathrm{BH}/M_*)^{1/3}$ is the tidal radius and $R_p$ is the pericenter distance).

 Recent works have shown that circularisation and initiation of accretion \emph{is} possible in the parabolic case, by a combination of energy dissipation in the `nozzle shock' that occurs on second pericenter passage (\citealt{steinbergstone24}; but see \citealt{bonnerotlu22} and Appendix~\ref{sec:resolution} for convergence studies of the nozzle shock) and/or relativistic precession leading to stream collisions \citep{andalmanetal22}.


In this paper, we present a set of simulations that self-consistently evolve a one solar mass polytropic star on a parabolic orbit around a $10^6$ solar mass black hole from the star's disruption to circularization of the returning debris and then accretion.
We follow the debris evolution for one year post-disruption, enabling us to approximately compute synthetic light curves which appear to match the key features of observations.

\begin{figure*}
   \centering
   \includegraphics[width=\textwidth]{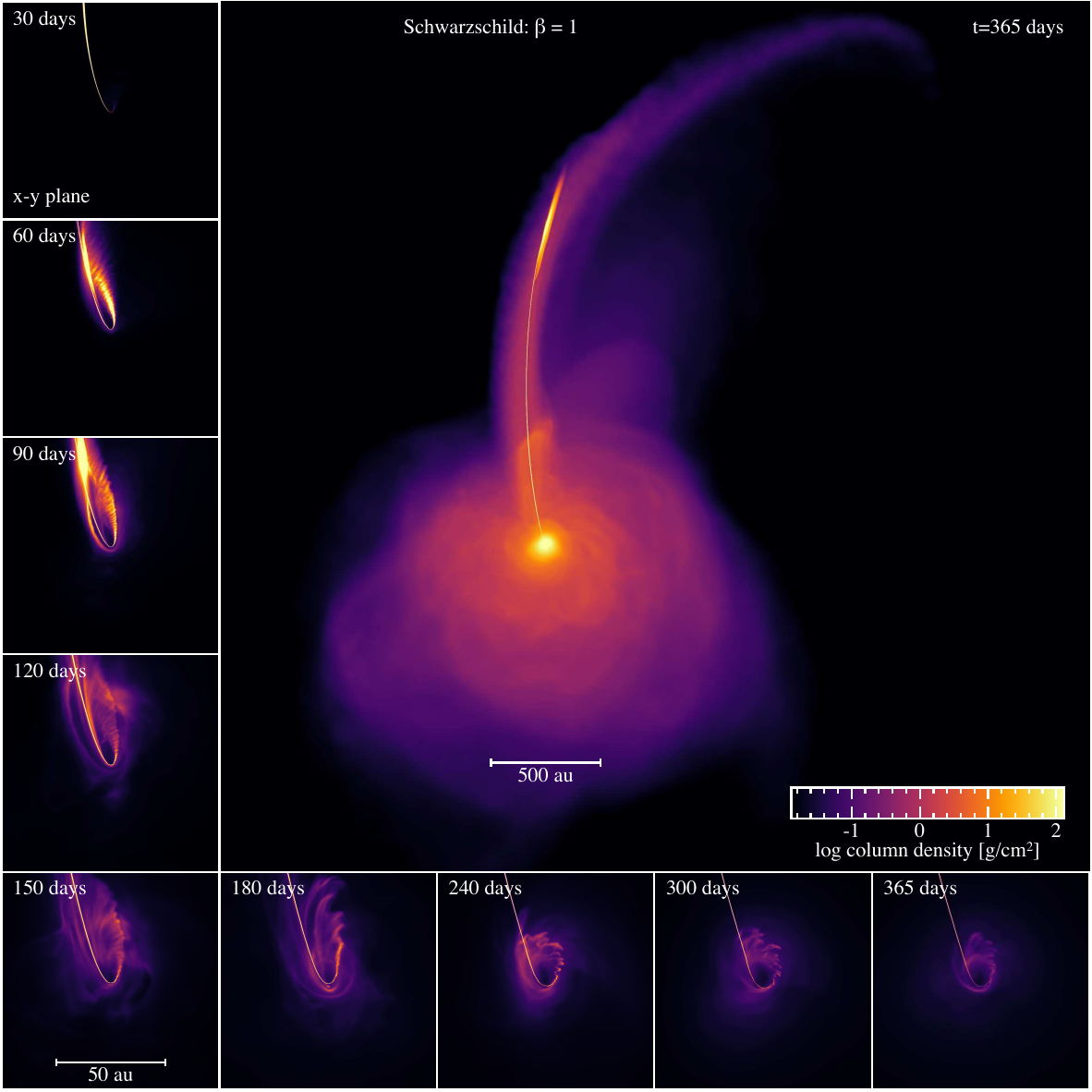}
   \caption{One year in the life of a tidal disruption event. We show shapshots of column density in the simulation of a $1M_\odot$ star on a parabolic orbit with $\beta=1$, disrupted by a $10^6M_\odot$ black hole, using $4 \times 10^6$ SPH particles in the Schwarzschild metric. \textit{Main} panel shows the large scale outflows after 365 days projected in the $x$-$y$ plane with log scale. Inset panels show the stream evolution on small scales (100$\times$100 au), showing snapshots of column density projected in the $x$-$y$ plane on a linear scale from 0 to 1500 g/cm$^2$ (colours are allowed to saturate). Animated versions of this figure are available in the online article.  Data and scripts used to create the figure are available on Zenodo:\dataset[doi:10.5281/zenodo.11438154]{https://doi.org/10.5281/zenodo.11438154}}
   \label{fig:main}
\end{figure*}

\section{Methods}

We modelled the disruption of stars with mass $M_*=1M_\odot$ and radius $R_*=1R_\odot$ by a supermassive black hole with mass $M_\mathrm{BH}=10^6M_\odot$ using the general relativistic implementation of the smoothed particle hydrodynamics code \textsc{Phantom} \citep{pricewurstertricco18} as described in \citet{liptaiprice19}.

We used the Kerr metric \citep{kerr63} in Boyer-Lindquist coordinates, with the black hole spin vector along the $z$-axis, and unless otherwise specified assumed an adiabatic equation of state for the gas with $\gamma = 5/3$.
We deleted particles that fall within a radius of $R_\mathrm{acc}=5R_g$ (within the innermost stable circular orbit of a non-spinning black hole) from the simulation to avoid small timesteps close to the horizon. 

We placed the star on a parabolic orbit initially at a distance of $r_0=10R_\mathrm{t}$ with velocity $v_0=\sqrt{2GM_\mathrm{BH}/r_0}$, where $R_\mathrm{t}$ is the tidal radius.
The parabolic trajectory was oriented such that the star travels counter-clockwise in the $x$-$y$ plane, and that the Newtonian pericentre $R_\mathrm{p}$ was located on the negative $y$ axis.
For inclined trajectories, we rotated the initial position and velocity vectors by an angle $\theta$ about the $y$-axis.

We modelled the star as a polytropic sphere with polytropic index $n=3/2$ using $4,188,898$ equal mass particles at our highest resolution and approximately $512$K particles in our medium resolution calculations (real stars are expected to be more concentrated, which may alter the fallback rate, delaying the peak with respect to a polytropic model; e.g. \citealt{Lawsmith19}), and include the perturbation in the metric due to Newtonian self-gravity in our simulations to hold the star together during the pre-disruption phase.

For our main simulations we employed an adiabatic approximation, implying that energy is trapped or advected rather than radiated. Appendix~\ref{sec:disc} reports a set of isentropic calculations where we assume shock heating to be immediately radiated away. The adiabatic simulations capture all accretion feedback down to $5R_{\rm g}$ inside which particles are assumed to accrete without radiating.

Our point of closest approach is typically smaller than the Newtonian pericentre $R_\mathrm{p}$ due to relativity, and thus the `true' penetration factor is larger (by up to $\sim\,$1 in our $\beta=5$ calculation, where the true $\beta$ is closer to 6).

\subsection{Synthetic lightcurves}
\label{sec:lightcurve-methods}


We computed synthetic lightcurves, optically thick visualisations, and spectra from our models by solving the equation of radiative transfer in the form
\begin{equation}
      \frac{{\rm d} I_\nu}{{\rm d} \tau} = S_\nu - I_\nu, \label{eq:rt}
\end{equation}
where $I_\nu \equiv I(\nu)$ is the intensity, $\tau$ is the optical depth and $\nu$ is the frequency. For simplicity we assumed local thermodynamic equilibrium such that $S_\nu \equiv B_\nu$, where $B_\nu = 2h\nu^3 c^{-2} \exp[h\nu/(k_{\rm B}T) - 1]^{-1}$ is the Planck function. We also neglected time-dependent and General Relativistic effects and special relativistic corrections to the shape of the photosphere \citep{abramowicz91}. That is, we considered rays to travel in a straight line to the observer. 
We solved for the optical depth using
\begin{equation}
{\rm d}\tau = \frac{\nu_0}{\nu} \kappa \rho  {\rm d} z' = \gamma (1 - \beta_z) \kappa \rho {\rm d}z'_0. \label{eq:dtau}
\end{equation}
where $\beta_z \equiv v_z/c$ and $\gamma = 1/\sqrt{1 - v^2/c^2}$ and $\rho$ is the density in the rest frame of the fluid. The correction factor $\nu_0/\nu \equiv \gamma (1 - \beta_z)$ accounts for the optical depth change caused by the moving photosphere \citep{abramowicz91,ogurafukue13}. We also computed $B_\nu$ in the observer's frame (i.e. using $\nu = \gamma (1 + \beta_z) \nu_0$), accounting for the Doppler shift due to the motion of the photosphere.

We found the relativistic corrections to the optical depth to be negligible in practice (because our typical $v/c \lesssim 1/30$), giving a 1-2\% correction to the total luminosity. This also justifies our neglect of photospheric shape distortions. Potentially more significant is the difference between the `thermalisation surface' and the photosphere in scattering-dominated flows \citep{abramowicz91,ogurafukue13}. Full treatment of this issue is beyond the scope of our paper (requiring a full 3D solution of the radiative transfer problem including scattering), but implies that we may underestimate the true temperature at the photosphere.

\begin{figure}
   \centering
   \includegraphics[width=\columnwidth]{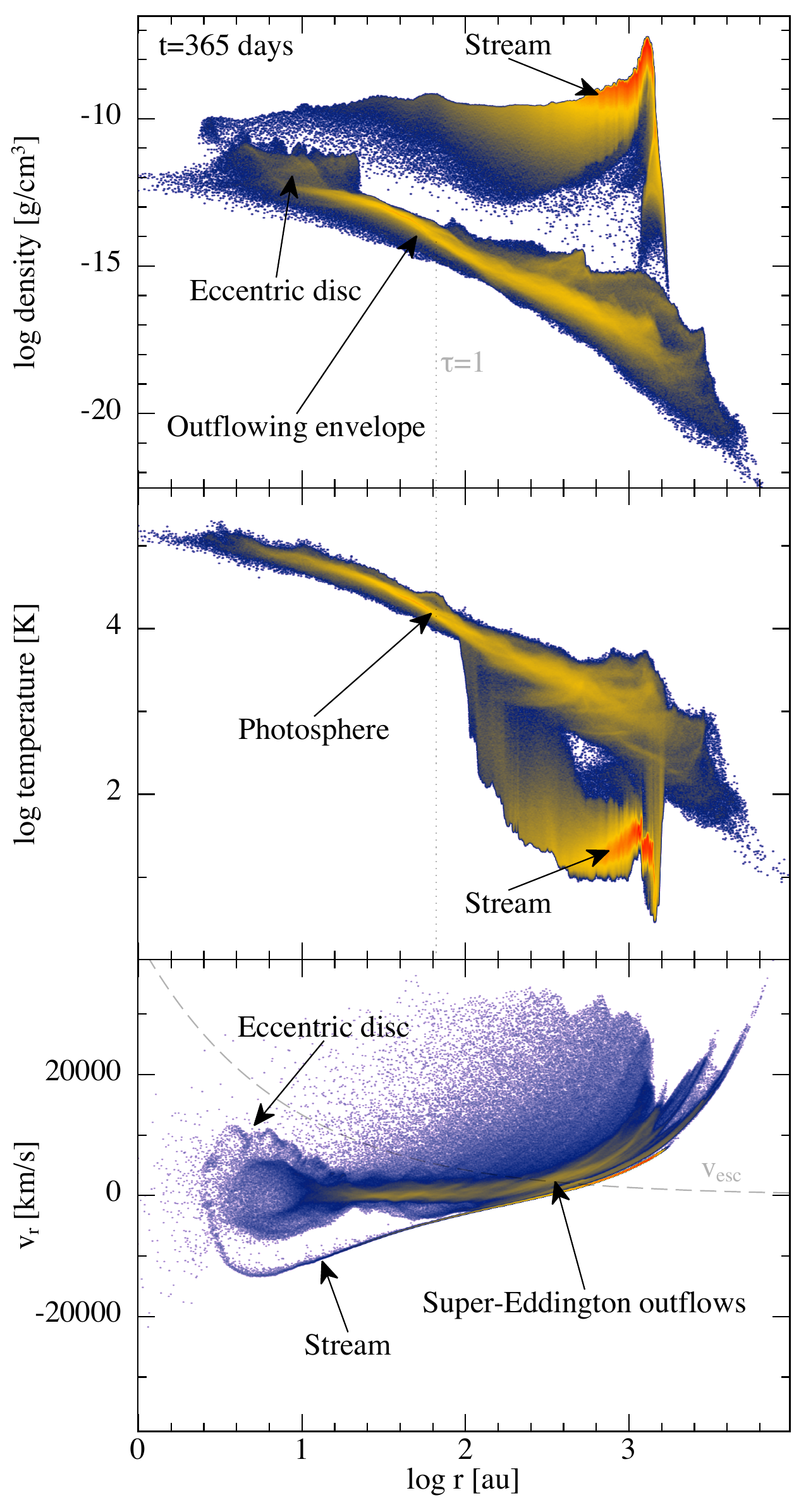}
   \caption{Distributions of density (\textit{top}), temperature (\textit{middle}) and radial velocity (\textit{bottom}) and  at one year post disruption. Colours represent the density of points (mass per pixel) on the plot, with orange being a high density of points and blue being low. Dotted line indicates the $\tau=1$ photosphere. Dashed line in the lower panel indicates $v_{\rm esc}\equiv \sqrt{2GM_{\rm BH}/r}$. All three distributions show features that may be identified as the high density stream and the low density, cool, expanding envelope. Typical outflow velocities are of order $10^4$ km s$^{-1}$, similar to those measured in observed TDE spectra and to the `mass loaded outflows' predicted by \citet{metzgerstone16}. An animated version of this plot is available online.  Data and scripts used to create the figure are available on Zenodo:\dataset[doi:10.5281/zenodo.11438154]{https://doi.org/10.5281/zenodo.11438154}}
   \label{fig:particle_plots}
\end{figure}




To compute the temperature from the specific internal energy, $u$, from the SPH calculation, we assumed gas and radiation to be locally in equilibrium, giving
\begin{equation}
f(T) =  \frac32 \frac{k_{\rm B} T}{\mu m_{\rm H}} + \frac{a T^4}{\rho} - u = 0,
\label{eq:tempfromu}
\end{equation}
where $\mu = 0.6$ is the mean molecular weight and $a$ is the radiation constant. We solved this for each particle via Newton-Raphson iteration until $\Delta T/T_0 < 10^{-8}$, where the initial guess $T_0 = \min[\frac23 u (\mu m_{\rm H}/k_{\rm B}),(\rho u/a)^{1/4}$].

 We adopted the opacity prescription from \citet[][see their figure 3 for a comparison to OPAL tables]{matsumotometzger22}, where
\begin{equation}
\kappa = \kappa_{\rm mol} + \kappa_e + (\kappa_{H^-}^{-1} + \kappa_{K}^{-1})^{-1}, \label{eq:opacitymm22}
\end{equation}
which includes molecular opacity $\kappa_{\rm mol} = 0.1 Z$ cm$^2$/g where $Z=0.014$ is the metallicity, $H^{-}$ opacity $\kappa_{H^-} = 1.1 \times 10^{-25} Z^{0.5} \rho^{0.5} T^{7.7}$ cm$^2$/g, bound-free and free-free opacity via Kramer's law $\kappa_{K} = 1.2 \times 10^{26} Z (1 + X) \rho T^{-7/2}$cm$^2$/g with $X=0.698$ (both formulae assuming $\rho$ in g/cm$^3$ and $T$ in K) and electron scattering
$\kappa_e = \sigma_e n_e/\rho$, where $\sigma_e = 6.652 \times 10^{-25}\,{\rm cm}^2$ is the Thomson cross section. We computed $n_e$ by solving for the hydrogen ionisation fraction, $x = n_e/n_H$ from the Saha equation (assuming H only for simplicity) by solving the quadratic
\begin{equation}
\frac{x^2}{1 - x} = \frac{1}{n_H} \left( \frac{2\pi m_e k_{\rm B} T}{h^2} \right)^{3/2} \exp\left[-\frac{13.6 {\rm eV}}{k_{\rm B}T}\right], \label{eq:saha}
\end{equation}
where $n_H = \rho/m_H$ is the hydrogen number density. In the limit of full ionisation this gives $\kappa_e = 0.4$ cm$^2$/g. The main effect of (\ref{eq:saha}) is to make regions with $T \lesssim 10^4$K transparent. In practice we found identical results with $\kappa = \kappa_e$ since electron scattering dominates the opacity at our typical photospheric densities ($\sim 10^{-14}$g/cm$^3$) and temperatures ($\sim 10^4$K).

We assumed a grid of pixels in the image plane with one ray per pixel, integrating back to front, each pixel containing 128 frequency bins spaced evenly in $\log_{10}\nu$ between $10^8$ and $10^{22}$ Hz. A caveat is that (\ref{eq:opacitymm22}) assumes grey opacities but we add blackbody spectra in a frequency-dependent manner. This is irrelevant anyway since the dominant opacity is grey.  Since our data consist of SPH particles, we performed the ray trace by sorting the particles by $z'$ (the line of sight towards the observer, not necessarily the original $z$), considering the propagation of the ray bundle in turn through each particle, computing the optical depth integral (\ref{eq:dtau}) through each particle using the smoothing kernel as described in \citet{price07}. We thus obtained a map of $I_\nu$ on our grid of pixels at 128 different frequencies. To `observe' the flow from different lines of sight, we simply rotated the particle distribution around the $x$, $y$ or $z$ axes while keeping the location of the observer fixed at $z' = \infty$ in the rotated coordinates ($x'$, $y'$ and $z'$). At early times the photosphere can be unresolved by the SPH particles \citep[e.g.][]{hatfulletal21}. To check for this we flagged pixels in our synthetic images where a single particle contributed more than ${\rm d}\tau > 1/3$ while $\tau < 1$. We then set a criterion on the unresolved area compared to the total area where $\tau > 1$. When $>1\%$ of pixels in the photosphere are poorly resolved we show the lightcurve using semi-transparent lines.

Synthetic spectral energy distributions were computed by integrating over all pixels in the image. Inferred bolometric luminosities were then computed by integrating the blackbody spectrum fitted to the optical band. To test our procedure we confirmed that imaging a $1 R_{\rm \odot}$ sphere of gas modelled with each particle given a uniform temperature $T = 5772$K and a constant opacity produced a luminosity equal to the solar luminosity.

\begin{figure}
   \centering
   \includegraphics[width=\columnwidth]{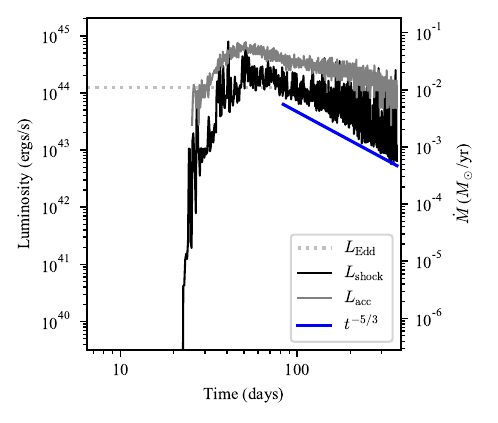}
   \includegraphics[width=\columnwidth]{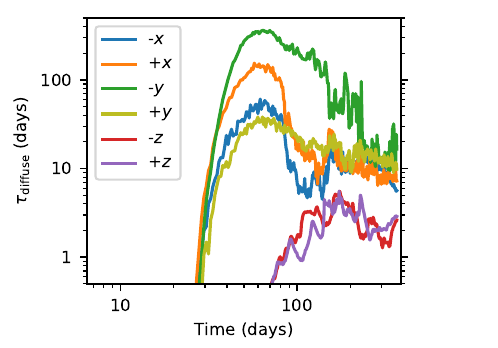}

   \caption{\textit{Top:} Energy dissipation rate and accretion luminosity evolution (from $\dot{M}$, see right axis), compared to the Eddington luminosity and a $t^{-5/3}$ slope. \textit{Bottom:} Estimated photon diffusion timescale along different lines of sight (see legend).  Data and scripts used to create the figure are available on Zenodo:\dataset[doi:10.5281/zenodo.11438154]{https://doi.org/10.5281/zenodo.11438154}}
   \label{fig:stuff_vs_time}
\end{figure}

\begin{figure*}
   \centering
   \includegraphics[width=\textwidth]{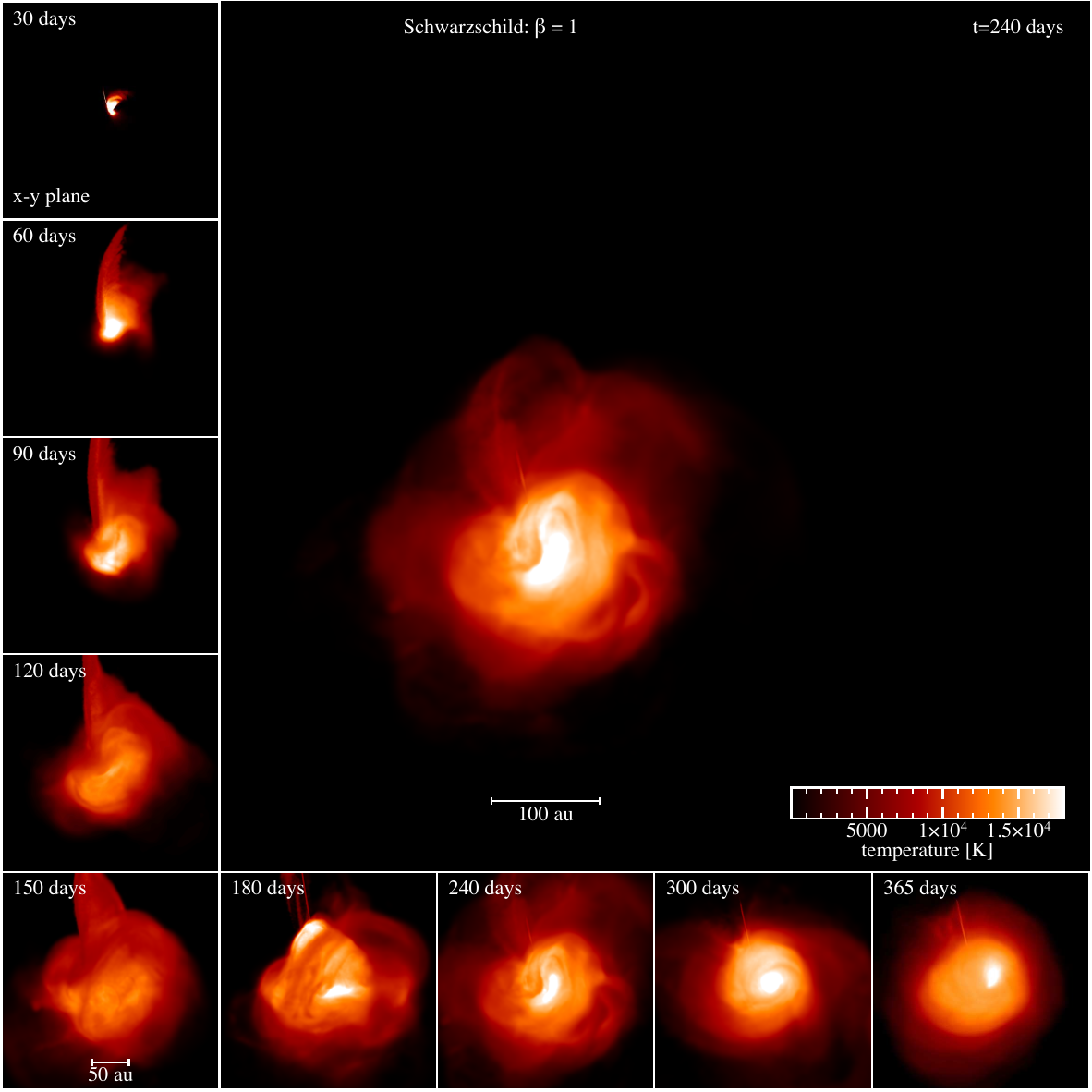}
   \caption{The Eddington envelope. We show temperature at the photosphere for the simulation snapshots shown in Figure~\ref{fig:main}. The inner engine is obscured by the optically thick envelope.  Temperatures correspond to thermal blackbody emission at UV/optical wavelengths.  Data and scripts used to create the figure are available on Zenodo:\dataset[doi:10.5281/zenodo.11438154]{https://doi.org/10.5281/zenodo.11438154}}
   \label{fig:photosphere}
\end{figure*}

\begin{figure*}
   \centering
   \includegraphics[width=\columnwidth]{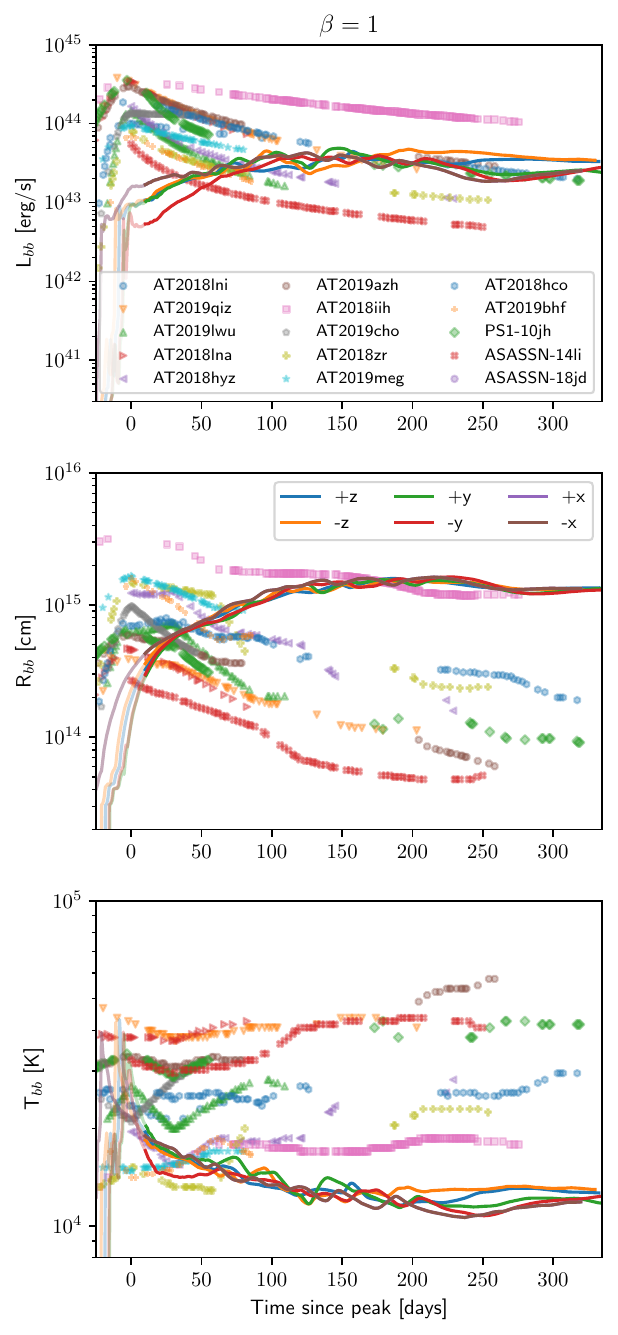}
      \includegraphics[width=\columnwidth]{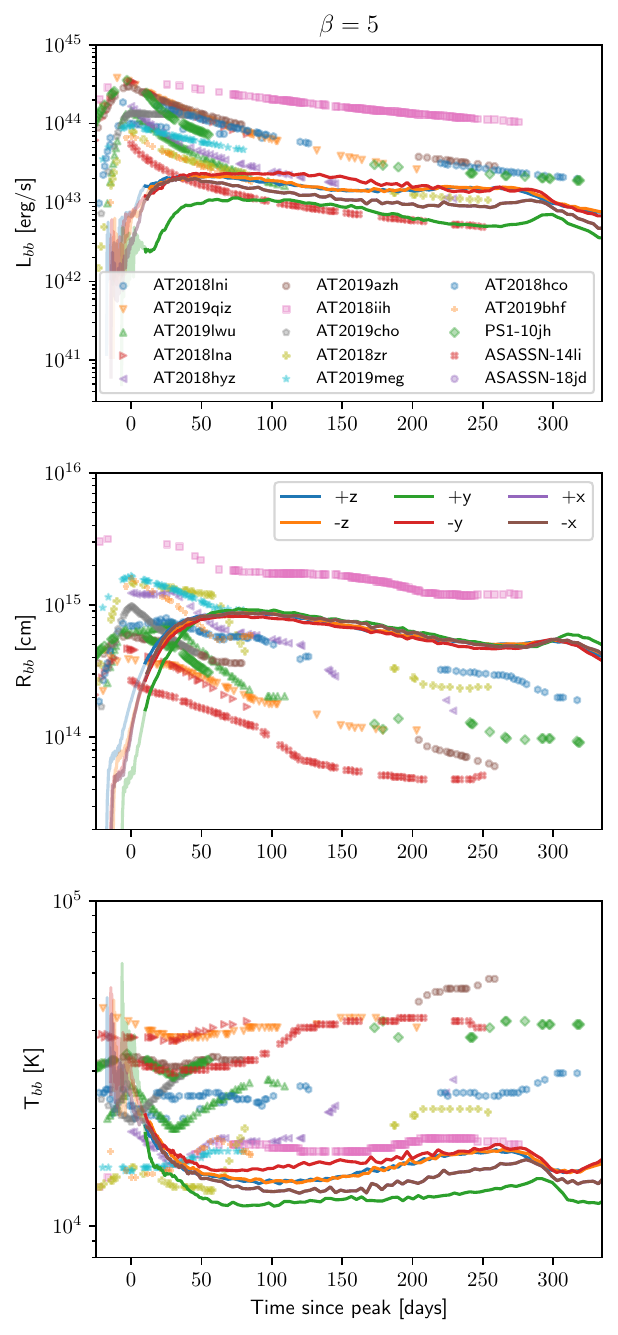}
   \caption{Optical luminosity (top), blackbody radius (middle) and temperature (bottom) evolution from our simulation using $\beta=1$ (left) and $\beta = 5$ (right), measured along different lines of sight (see legend). Markers represent the fits to various observed TDEs taken from \citet{van-velzenetal21} plus ASASSN-18jd \citep{neustadt20}. Transparency at $t \lesssim 30$ days indicates when the photosphere is unresolved.  Data and scripts used to create the figure are available on Zenodo:\dataset[doi:10.5281/zenodo.11438154]{https://doi.org/10.5281/zenodo.11438154}}

   \label{fig:lightcurves}
\end{figure*}

\section{Results}

Figure~\ref{fig:main} shows snapshots of the long-term debris evolution in the $\beta=1$, high-resolution, adiabatic simulation with a non-rotating black hole.
The star takes approximately 7 hours to reach pericentre at which point it is tidally disrupted and forms a long, thin debris stream.
Approximately half of the stream is bound to the black hole while the other half is unbound (visible as the thin, dense upward stream in the figure).
The head of the stream (i.e., the most bound material) takes $\sim\,$26 days to return to pericentre.
After the stream passes through pericentre for a second time, the small amount of apsidal precession leads to stream self-intersection at the `outer shock' (most evident in inset panels shown at 60 and 90 days; see Appendix~\ref{sec:fanning} for more detail on the outer shock), which cause material to eventually fall towards the black hole, driving the formation of kinetic outflows which spread almost isotropically, eventually encasing the black hole (main panel).
A fraction of the material joins the returning debris and returns to pericentre, however most of it goes into a low density mechanical outflow.
Following the fallback for one year post disruption, we find that the bubble continues to expand, with only a small amount of material forming an eccentric disc (lower panels of Figure~\ref{fig:main}). The large-scale debris structure (main panel) is similar to the `Eddington envelope' first proposed by \citet{loebulmer97} except that ours is outflowing rather than static, as predicted by subsequent authors \citep{strubbequataert09,jiangguillochonloeb16,metzgerstone16}.

Figure~\ref{fig:particle_plots} shows the density (\textit{top}), temperature (\textit{middle}) and radial velocity (\textit{bottom}) distributions of particles a year after disruption, with temperature calculated from Equation~(\ref{eq:tempfromu}).
The density of the outflowing envelope is $\sim\,$$10^{-16}$ g cm$^{-3}$, approximately six orders of magnitude lower than the density of the stream, which is $\sim\,$$10^{-10}$ g cm$^{-3}$.
The typical outflow velocity is $\sim\,$10$^4$ km s$^{-1}$. After one year there is approximately 67\% ($0.67$~M$_\odot$) of the stellar mass in the incoming or unbound stream, 31.2\% ($0.31$~M$_\odot$) in the envelope, 0.8\% ($0.008$~M$_\odot$) in the eccentric disc, while 1\% ($0.01$~M$_\odot$) has been accreted by the black hole.

The top panel of Figure~\ref{fig:stuff_vs_time} shows the energy dissipation rate in our simulation, $L_\mathrm{shock}$.
This was computed by computing the time derivative of the thermal energy caused by irreversible heating terms due to viscosity and shocks. We find that it is dominated by regions close the accretion radius $R_\mathrm{acc}$.
The energy dissipation rate rises rapidly from $\sim\,$26 days, when accretion begins, until $\sim\,$50 days after disruption when it reaches a peak of $\sim\,$$10^{44}$ erg s$^{-1}$ (roughly the Eddington limit).
This is followed by a power law decay that does approximately match the expected $t^{-5/3}$ energy dissipation from the mass fallback rate \citep{rees88,phinney89}, but with $\dot{M}_{\rm acc}$ lower than the mass fallback rate by approximately two orders of magnitude.
We compare this to the luminosity derived from our measured mass accretion rate $\dot{M}$, given by $L_\mathrm{acc}=\epsilon \dot{M} c^2$.
Here, $\epsilon=0.2$ is the efficiency based on our inner accretion radius. However, since our $R_{\rm acc}$ is inside the last stable orbit, matter crossing this radius is mostly advected into the black hole rather than radiated, so converting it into a luminosity simply measures the amount of energy that would have been dissipated to reach this radius if all the matter were on a perfectly circular orbit.
The resulting `luminosity' $L_\mathrm{acc}$ follows the same trend as the actual energy dissipation rate above, but is $\sim$0.5--1 orders of magnitude greater throughout the simulation.
This suggests some material is accreted almost radially, with little to no orbital energy dissipated before accretion \citep{svirskipirankrolik17}.

The lower panel of Figure~\ref{fig:stuff_vs_time} shows the approximate time taken for photons to diffuse from near the black hole out to the photosphere radius, assuming the material remains static.
We computed this via
\begin{equation}
   \tau_\mathrm{diffuse} \sim \frac{1}{c} \int_0^{R_\mathrm{photo}} \kappa\,r\,\rho\,dr,
\end{equation}
by summing the contributions from each SPH particle intersecting a line-of-sight along each coordinate axis with opacity $\kappa = 0.4$ cm$^2$/g.
The typical photon diffusion time varies between $\sim\,$10--200 days, peaking approximately 80 days after disruption and decaying roughly as a power law. This indicates that, although most energy is dissipated in regions close to the black hole (see Figure~\ref{fig:thermo}), where the reservoir of gravitational binding energy is largest, this material is optically thick.
The energy is transported outward by mechanical outflows before some of it is radiated through a photosphere.

Figure~\ref{fig:photosphere} shows the temperature at the `photosphere' of the outflowing envelope, produced by ray-tracing through our simulation as described in Sec.~\ref{sec:lightcurve-methods}. An expanding emission region of 10--100 au is consistent with those inferred from TDE observations \citep{van-velzenetal21}. Photospheric temperatures are in the range expected to produce blackbody emission at UV/optical wavelength. Figure~\ref{fig:particle_plots} shows that temperatures would remain in the 10$^4$--10$^{5}$ K range even accounting for scattering from deeper layers (see Appendix~\ref{sec:lightcurves}).

Figure~\ref{fig:lightcurves} compares the corresponding time evolution of our measured $L_\mathrm{bb}$, $R_\mathrm{photo}$ and $T_\mathrm{bb}$ (inferred bolometric luminosity, blackbody radius and temperature of the blackbody fit to optical wavelengths, respectively) from six different lines of sight (solid lines) to the evolution of observed TDEs inferred from black body fits taken from \citet{van-velzenetal21} and \citet{neustadt20} (dots). The equivalent plot for the more penetrating $\beta = 5$ encounter is shown on the right.

The typical luminosity (top panel) is around $10^{43}$--$\,10^{44}$ erg s$^{-1}$ ($10^{9.4}$--$\,10^{10.4}$ $L_\odot$), however we caution that the exact value is sensitive to our photospheric temperature (since $L\propto T^4$). Nonetheless, our approximate luminosities are consistent with the observed values (see figure 9 of \citealt{holoienhubershappee19}), as well as with the analytically computed light curves of \citet{lodatorossi11} (see left panel of their Figure 7). The bolometric luminosity of the envelope in our models appears almost constant, and close to the Eddington value, key features of the analytical model of \citet{loebulmer97}.

The photospheric radius (centre panels), when computed by fitting to synthetic spectra in the optical band (Appendix~\ref{sec:lightcurves}), is relatively insensitive to the line-of-sight, giving typical radii of $\sim\,$10--100 au ($10^{14}$--$10^{15}$ cm), in the same range as the observations. The photospheric radius  reaches a peak after roughly 150--200 days in the $\beta = 1$ calculation, and $50$ days in the $\beta = 5$ calculation, the latter more closely matching the typical $20-50$ day rise time seen in most of the observations. On the other hand some TDE candidates such as ASASSN-18jd \citep{neustadt20} (see legend) do show a more or less flat luminosity evolution over a 1 yr timescale, similar to our $\beta=1$ lightcurves. The transients AT2018hco and AT2019iih show the closest match to the photospheric radius evolution in our $\beta=5$ calculation, particularly for $t \gtrsim 50$ days. 


We find photospheric temperatures (lower panels) in the range of $1$--$4 \times 10^4 K$, with the same range as the observations, but with slightly lower temperatures in our models ($T \sim 10^{4.1}$K) compared to the observations ($T\sim 10^{4.2}$--$10^{4.6}$K) for $t \gtrsim 50$ days. This may be a result of our underestimate of the photospheric temperature due to scattering of photons from deeper layers (Appendix~\ref{sec:lightcurves}). The observed trend of temperatures slowly rising as a function of time for $t \gtrsim 50$ days is reproduced in our $\beta=5$ calculation (right panels of Figure~\ref{fig:lightcurves}). Our temperature evolution is similar to that obtained by \citet{mocklerguillochonramirez-ruiz19} who fitted models of a `dynamic reprocessing layer' to observed TDEs (see their Figure 8), finding a 10--20 day sharp rise in temperature followed by a dip and subsequent flat or slowly rising evolution.




A notable feature in both the $\beta=1$ and $\beta=5$ cases are the almost flat lightcurves predicted at late times along most lines of sight. This is also observed \citep[e.g.][]{holoienhubershappee19,van-velzenetal21}. In our simulations this corresponds to the stalling of the photospheric radii in an expanding gas cloud; however, as discussed below, models without cooling may not accurately describe light curves after $\sim 1$ year.

\section*{Discussion}
Our simulations address the main challenge of tidal disruption events: to simulate the disruption of a $1M_\odot$ star, on a parabolic orbit, by a $10^6 M_\odot$ black hole in general relativity, and to model the subsequent accretion flow from first principles.

Our main new insight from doing so is that, as in \citet{huetal24}, we have been able to self-consistently form the hypothesised `Eddington envelope' for the first time, more or less as predicted by \citet{loebulmer97} and \citet{metzgerstone16}. With the disrupted star moving on an orbit with a mean energy of zero and low angular momentum, we found dissipation of energy and angular momentum to occur through relativistic precession and ensuing stream-stream collisions (see Appendix~\ref{sec:disc}), as found by previous authors for stars on bound trajectories \citep{hayasakistoneloeb16,bonnerotrossilodato16,stonekesdencheng19,liptaipricemandel19} and similar to the process originally envisioned by \citet{rees88}.

Once accretion commences in the optically thick environment with inefficient cooling (we assumed adiabatic gas), the gravitational energy of accreting material launches mechanical outflows, yielding a large, asymmetric, optically thick outflowing envelope (Fig.~\ref{fig:photosphere}). 

 We found a similar outflow in recent calculations of initially eccentric encounters \citep{huetal24}, but with faster rise of the lightcurve due to the shorter orbital periods. The main difference in the parabolic case is that only $\sim 1\%$ of the star is accreted, compared to more than half the star for the eccentric encounters \citep{huetal24}. The lower efficiency of accretion occurs because of the smaller amount of bound material reaching the inner regions (Appendix~\ref{sec:disc}).

The envelope structure is similar to other rapidly expanding, optically thick, quasi-spherical outflows powered by a central energy source, such as supernovae and kilonovae.
In this case, the expanding envelope is powered by gravitational energy, dissipated in the radiatively inefficient accretion flow onto the black hole (see Figure~\ref{fig:thermo}). Our black hole mass accretion rate, shown in Figure~\ref{fig:stuff_vs_time}, is 1--1.5 orders of magnitude lower than the predicted mass fallback rate.
This broad conclusion on the formation of an Eddington envelope is unchanged by whether or not the black hole is spinning, or by whether the star approaches closer to the black hole (see Figure~\ref{fig:beta5}), or the numerical resolution (see Appendix~\ref{sec:resolution}).

That our outflowing envelope, or `Black Hole Sun', is optically thick is demonstrated by our estimated photon diffusion timescales of $\sim\,$10--100 days shown in the bottom panel of Figure~\ref{fig:stuff_vs_time}.
This means that the debris produced by the disruption cannot cool efficiently, leading to the formation of a photosphere. Synthetic spectral energy distributions (Appendix~\ref{sec:lightcurves}) show this leads to ratios of optical to X-ray emission of order $10$--$10^3$, similar to those observed \citep{van-velzenetal21}.


Our numerical experiments have several serious limitations in how well they represent real tidal disruptions. The main ones are that we employed an adiabatic approximation, meaning that none of the energy produced is actually radiated in our simulations, that we assume $\gamma = 5/3$ and used a polytrope not a real star.

Our adiabatic assumption is likely a good approximation near the peak of the light curve since the photon diffusion timescale is long, although it does not account for the change to the radiation-pressure dominated regime after shock heating, which implies that the adiabatic index should change to $4/3$.
On the other hand, the adiabatic assumption likely breaks down by around 1 year after disruption.
At t $\gtrsim$ 1 yr, our estimated luminosity (Figure~\ref{fig:lightcurves}) becomes comparable to the energy input rate from shock heating (Figure \ref{fig:stuff_vs_time}), while the photon diffusion timescale becomes comparable to the time for the kinetic outflows to reach the photosphere, allowing material to radiatively cool.

Therefore, at t $\gtrsim$ 1 yr, with relatively efficient diffusion, the observable luminosity should transition to tracking the shock heating rate, which in turn roughly tracks the mass return rate with $\propto t^{-5/3}$ scaling (top panel of Figure \ref{fig:stuff_vs_time}). Efficient cooling after 1 year may also lead to thin disc formation (see Appendix~\ref{sec:disc}), rather than the eccentric thick disc formed at $t \gtrsim 240$ days in Figure~\ref{fig:main}. Cooling may also cause our photospheres to retreat at later times, better matching the photospheric radius and temperature  observations.

Despite this limitation, our photospheric radius estimates yield a luminosity evolution with the same order of magnitude as optical/UV light curves from several recently observed TDEs (see Figure~\ref{fig:lightcurves}).
We do not expect that the dynamics of our simulations would be significantly affected by allowing the photosphere to radiate at early times, mainly because the energy release via radiation is small --- most of the energy in our Eddington envelope is released via mechanical outflows, with typical outflow velocities of $\sim\,$1--2$\times 10^4$ km s$^{-1}$ (see bottom panel of Figure~\ref{fig:particle_plots}).
Such velocities match those observed in spectral lines \citep[e.g.][]{leloudasdaiarcavi19}.

We find that the optical light curve is sensitive to the observing angle, as suggested by \citet{daimckinneyroth18}, but due to different photospheric temperatures along different lines of sight (see Figure~\ref{fig:lightcurves}) rather than the presence of a disc.
Mildly supersonic outflows indicate that temperatures are not in equilibrium across the photosphere.

The photospheric radii plateau at $\sim 20$ to $\sim 100$ au, depending on the line of sight and the amount of material participating in the outflow (Figure~\ref{fig:lightcurves}). Applying a single-zone model with uniform density $\rho = \frac{4\pi}{3} \frac{M}{R^3}$ to a homologously expanding envelope, this stalling is estimated to occur when the photospheric radius is $\frac{2}{3}\sqrt{\frac{\kappa M}{4\pi}}$, where the mass $M$ in the envelope depends on the penetration factor $\beta$.  This estimate yields a photospheric radius plateau of $\sim 100$ au for our choice of $\kappa$ and $M=0.1 M_\odot$.
While our predicted radius evolution might change if we allowed radiation to escape (especially at these late times), a significant late-time flattening has been observed in optical/UV for black holes of this mass \citep{van-velzenstonemetzger19}.

Many TDE candidates show soft X-ray emission. While in general the ratio of X-ray to optical luminosities in TDE is low \citep{auchettlguillochonramirez-ruiz17}, there are cases where optical and X-ray emission are both present.  ASASSN-14li has similar X-ray and optical luminosities \citep{holoienkochanekprieto16}, though with X-rays lagging by 32 days \citep{pashamcenkosadowski17}, which is difficult to explain via reprocessing. Our model can help to explain some of these TDE features. At late times, as the material cools and the photosphere retreats to reveal regions close to the black hole, we expect the formation of a disc or a thick torus similar to that described in Appendix~\ref{sec:disc}, which will be a significant central X-ray source.

Indeed, we found in our $\beta = 5$ calculation (Appendix~\ref{sec:lightcurves}) that retreat of the photosphere produces a rising X-ray luminosity similar to some observed TDEs \citet{gezaricenkoarcavi17} and to the class of `veiled' TDEs \citep{auchettlguillochonramirez-ruiz17}. Our Eddington envelope naturally obscures any central X-ray source, with absorption strongly varying for different lines of sight with typical $N_H \sim 10^{24}$--$10^{26}$ cm$^{-2}$. This suggests an explanation for the dichotomy between X-ray and optical TDE in terms of different viewing angles to the source, in a sort of `unified' model for TDE emission \citep{Dai18}.

\section{Conclusions}
In summary, we simulated the tidal disruption of 1M$_\odot$ polytropic stars around a $10^6$ solar mass black hole, using general relativistic hydrodynamics in the Kerr metric, following the simulations for one year post disruption. Our main finding, similar to our recent finding for real stars on eccentric orbits \citep{huetal24} is the production of an optically thick `Eddington envelope' hypothesised by previous authors \citep[e.g.][]{loebulmer97,metzgerstone16}. Our calculations predict:
\begin{enumerate}
   \item outflow velocities of $\sim\,$$10^4$ km s$^{-1}$;
   \item large photosphere radii of $\sim\,$$10$--$100$ au;
   \item a relatively low mass accretion rate ($\sim\,$$10^{-2} M_\odot/$yr) mediated by the mechanical outflow of material;
   \item peak luminosities of $\sim\,$$10^{43}$--$10^{44}$ erg/s.
\end{enumerate}
We find that our $\beta = 5$ encounters  provide a better match to TDE observations than $\beta = 1$, and that black hole spin does not significantly change the outcome. We thus confirm Eddington envelopes as the likely solution to the puzzle of TDE optical/UV emission.


\begin{acknowledgments}
We thank the referee for constructive suggestions and Christophe Pinte, Benjamin Tessore, Sjoert Van Velzen, Nick Stone, Brian Metzger, Katie Auchettl, Matt Nicholl, James Miller-Jones, Adelle Goodwin, Jane Dai, Kate Alexander, Paul Lasky, Elli Borchert, Alex Heger, Brenna Mockler, Selma de Mink, Nathaniel Roth, Greg Salvesen, Eliot Quataert and Bernhard Mueller for stimulating discussions. Thanks to Sjoert and Katie in particular for kindly sharing observational data.
DL was funded through the Australian government Research Training Program.
DJP and IM are grateful for Australian Research Council funding via FT130100034 and FT190100574, respectively.
This project received funding from the EU's Horizon 2020 research and innovation programme under the Marie Sklodowska-Curie grant 823823 (Dustbusters RISE project). Research at Flatiron Institute was supported by the Simons Foundation. DJP thanks them for expenses during his visit. This research was supported by grant NSF PHY-2309135 to the Kavli Institute for Theoretical Physics (KITP).
\end{acknowledgments}

%

\vspace{5mm}
\facilities{We acknowledge CPU time on OzSTAR funded by Swinburne University and the Australian Government.}


\software{\textsc{phantom} \citep{pricewurstertricco18},  
          \textsc{splash} \citep{price07}, \textsc{matplotlib}, \textsc{scipy}
          }



\bibliography{dave}{}
\bibliographystyle{aasjournal}

\appendix

\section{Radiatively efficient cooling and disc formation}
\label{sec:disc}

\begin{figure*}
   \centering
   \includegraphics[width=\textwidth]{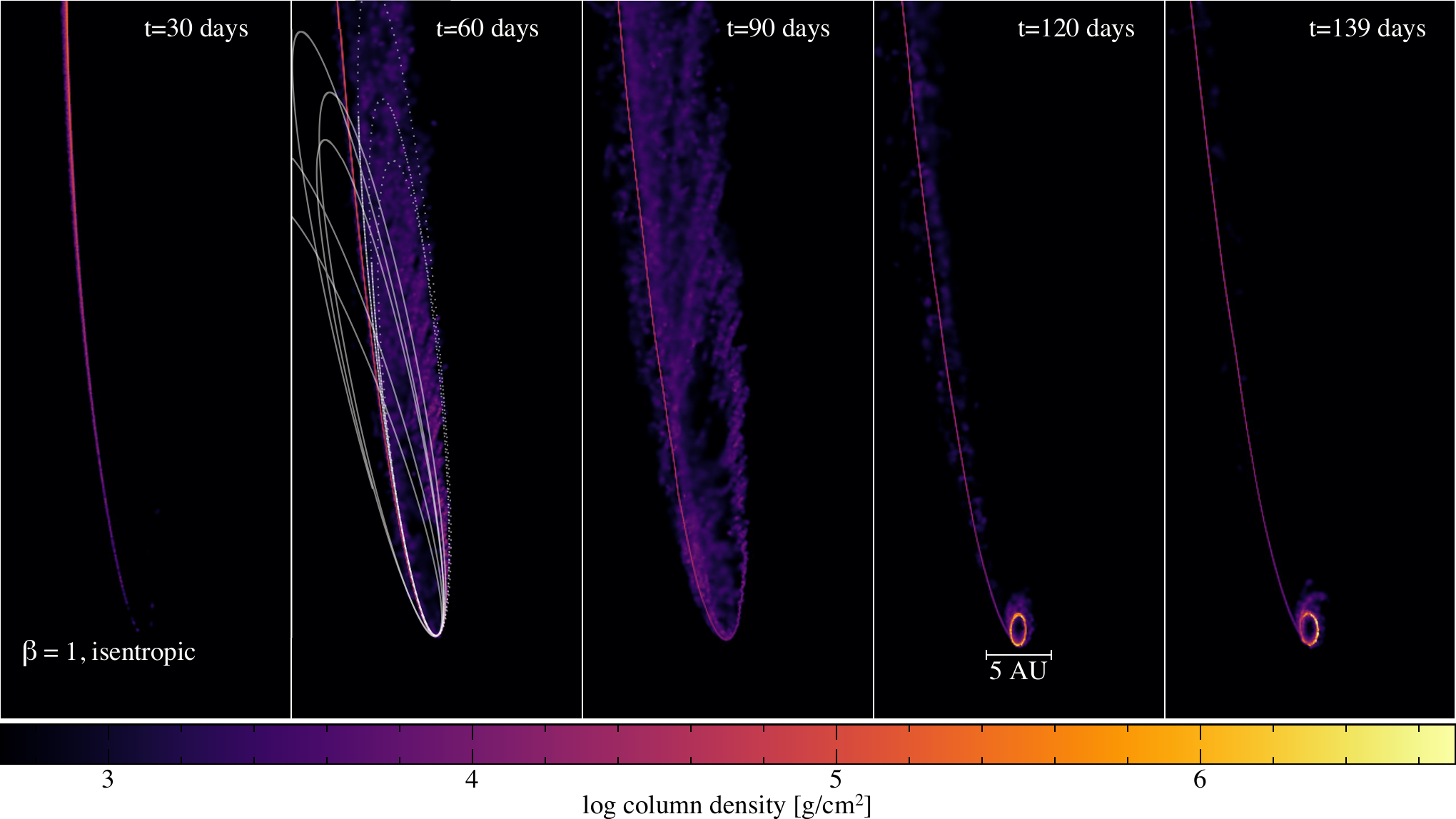}
   \caption{Disc formation in an isentropic simulation (radiatively efficient cooling) with a non-spinning black hole, as in Fig.~\ref{fig:main}. After $\sim 120$ days a narrow, circular, ring of material forms in the orbital plane. On the second panel we show ballistic trajectories of three selected particles in the debris stream (solid white lines), showing how relativistic orbital precession leads to repeated self-intersection of the stream and subsequent disc formation. Dashed white lines show the corresponding Newtonian trajectories (i.e. without relativistic precession). View is inclined by $60^\circ$ to match Fig.~\ref{fig:disc_formation_kerr}. Data and scripts used to create the figure are available on Zenodo:\dataset[doi:10.5281/zenodo.11438154]{https://doi.org/10.5281/zenodo.11438154}}
   \label{fig:disc_formation}
\end{figure*}

\begin{figure*}
   \centering
   \includegraphics[width=\textwidth]{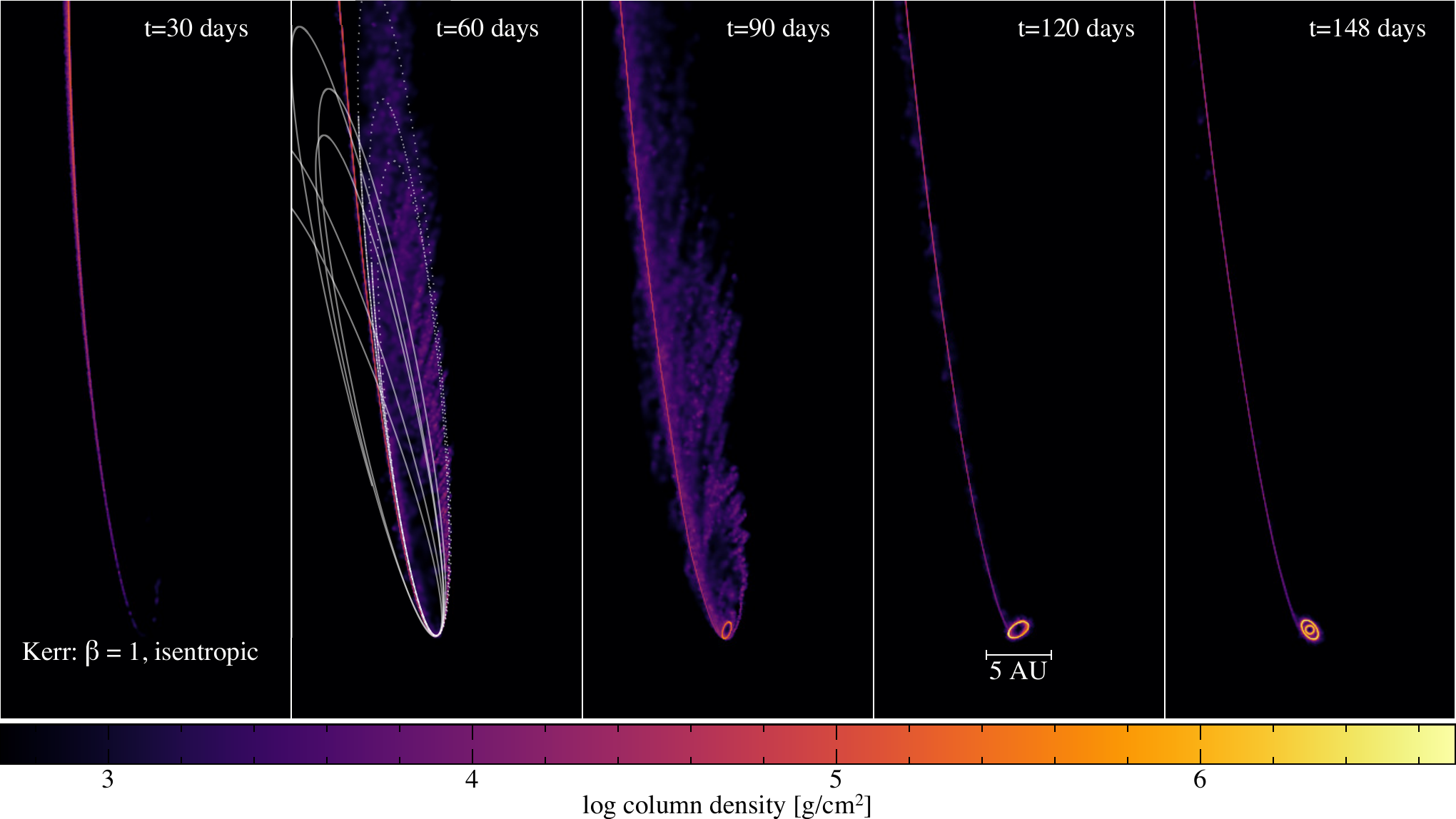}
   \caption{Disc formation in an isentropic simulation (radiatively efficient cooling) with a spinning black hole ($a=0.99$, $\theta = 60^\circ$). The disc in this case is formed after 90 days, and undergoes differential precession which `tears' the disc into independent rings. Data and scripts used to create the figure are available on Zenodo:\dataset[doi:10.5281/zenodo.11438154]{https://doi.org/10.5281/zenodo.11438154}}
   \label{fig:disc_formation_kerr}
\end{figure*}

Figure~\ref{fig:disc_formation} demonstrates that disc formation can occur more rapidly if the gas is allowed to cool. We performed a medium-resolution simulation with the same parameters as in Fig.~\ref{fig:main} but using an isentropic approximation in which we discard all irreversible heating due to viscosity and shocks, but retain heating and cooling due to $P{\rm d}V$ work; see \citealt{liptaipricemandel19}). This shows that a compact, circularised disc can more efficiently form if the stream remains cold. The
formation process is similar to the stream-stream collisions seen in simulations of tidally disrupted stars on bound orbits \citep{hayasakistoneloeb16,bonnerotrossilodato16a} caused by the general relativistic orbital precession \citep{rees88}, except that the collisions occur at relatively large distances from the black hole ($\sim 10$s of au) for our $\beta = 1$ encounter. No optically thick envelope is formed in the isentropic case as the main energy source is switched off.

The second panel of Figure~\ref{fig:disc_formation} quantifies the degree of relativistic apsidal precession, plotting three sample trajectories computed by integrating the geodesic equations\footnote{\url{https://github.com/phantomSPH/phantom-geodesic}} for three sample particles in the returning debris stream with $a=0$ (solid lines), which may be compared to the corresponding Newtonian trajectories (dashed lines). The relativistic precession, while small for $\beta = 1$, nevertheless leads to self-intersection of the stream and hence disc formation.

Figure~\ref{fig:disc_formation_kerr} shows the same calculation but performed with a spinning black hole ($a=0.99$; $\theta = 60^\circ$). The disc formation process can be seen to occur in a similar manner. The second panel shows the same set of orbital trajectories as in Figure~\ref{fig:disc_formation}, but computed with $a=0.99$. These are indistinguishable from those shown in Figure~\ref{fig:disc_formation}, indicating that the spin does not play a large role in the disc formation process for $\beta=1$ encounters. The main difference is that in the case of a spinning black hole, the disc that is subsequently formed undergoes differential Lense-Thirring precession leading to disc tearing as predicted by \citet{nixonkingprice12} and \citet{nealonpricenixon15}. Such differentially precessing rings are thought to be responsible for Quasi-Periodic Oscillations in black hole accretion discs \citep{stellavietri98} and are also observed in tidal disruption events \citep{pashametal19}.

This approximation of efficient cooling throughout the disruption event is unrealistic. Hence we also performed an additional numerical experiment where we continued the medium-resolution adiabatic calculation (with the same parameters as Fig.~\ref{fig:main}) but at 1 year after disruption switched to an isentropic formulation. This corresponds to roughly the time when the photon diffusion time becomes short (Fig.~\ref{fig:stuff_vs_time} \textit{bottom}) and the adiabatic cooling rate exceeds the energy dissipation rate. However, in reality, the envelope would transition smoothly from nearly adiabatic expansion with inefficient cooling to efficient radiative cooling represented by the isentropic model. That we also form a circularised disc in this experiment confirms that a disc would efficiently form even at low resolution if material was allowed to cool.


\begin{figure*}
   \centering
   \includegraphics[width=\textwidth]{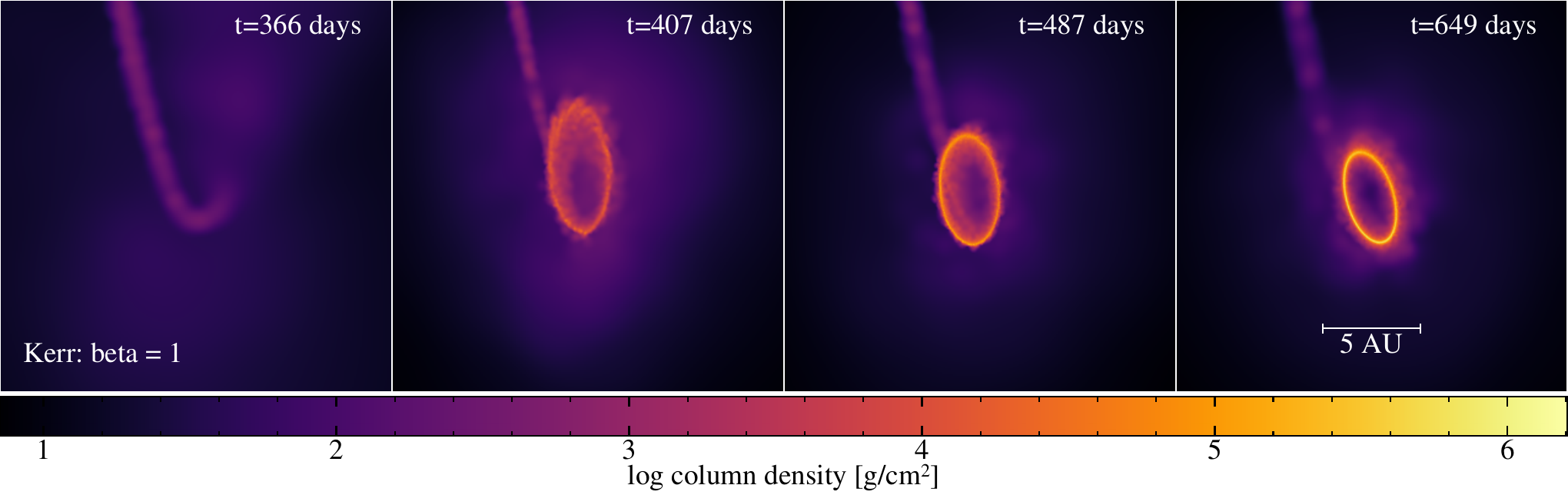}
   \caption{Disc formation in the core of the envelope when an adiabatic simulation is continued with radiatively efficient cooling. We resumed an adiabatic calculation (as in Figure~\ref{fig:main}) one year after disruption assuming isentropic evolution. A narrow ring of material develops, similar to that obtained in the fully isentropic calculation (Figure~\ref{fig:disc_formation}). Each panel is 20 au $\times$ 20 au.}
   \label{fig:disc_formation2}
\end{figure*}

\section{Thermodynamics}\label{sec:fanning}
\begin{figure*}
   \centering
   \includegraphics[width=0.49\textwidth]{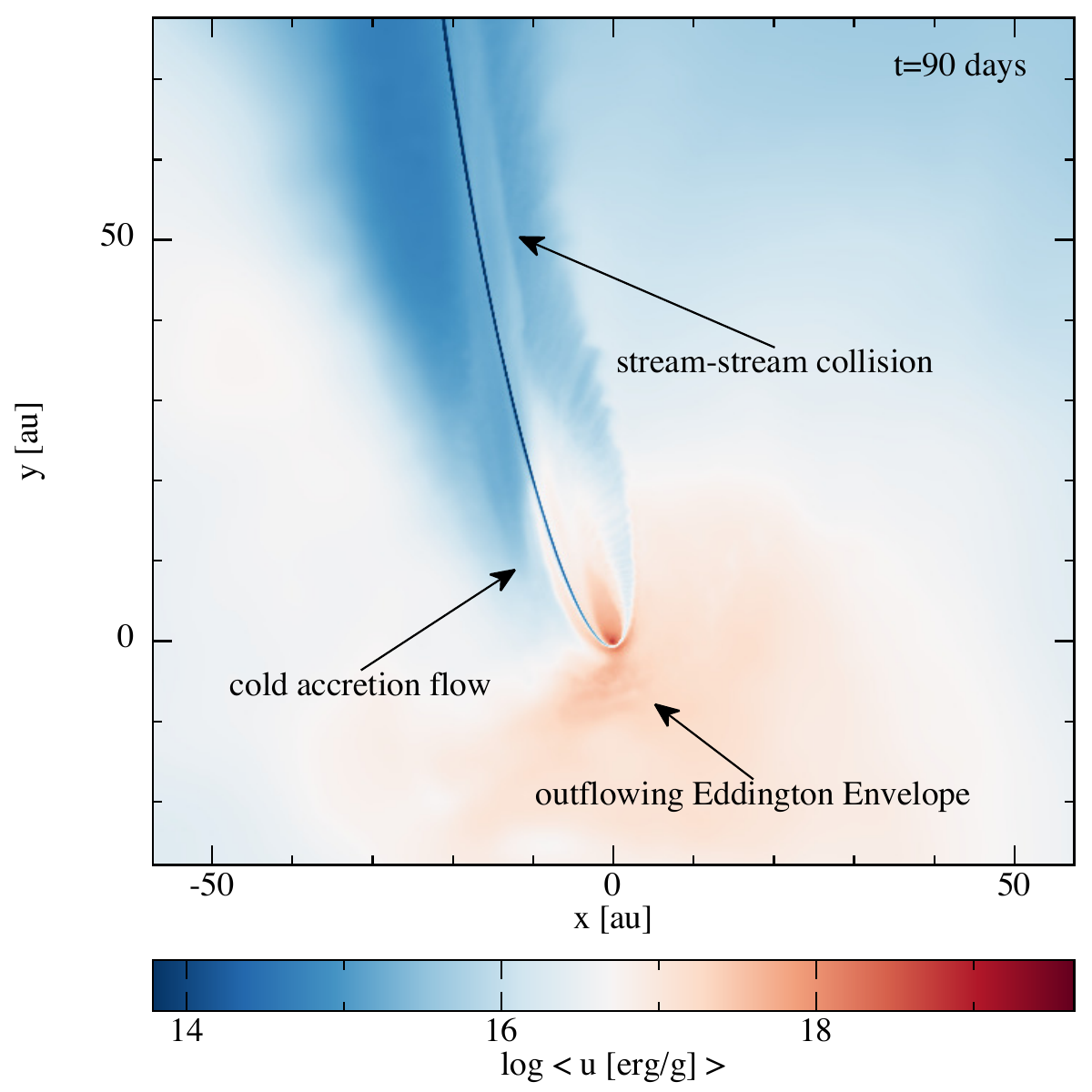}
      \includegraphics[width=0.49\textwidth]{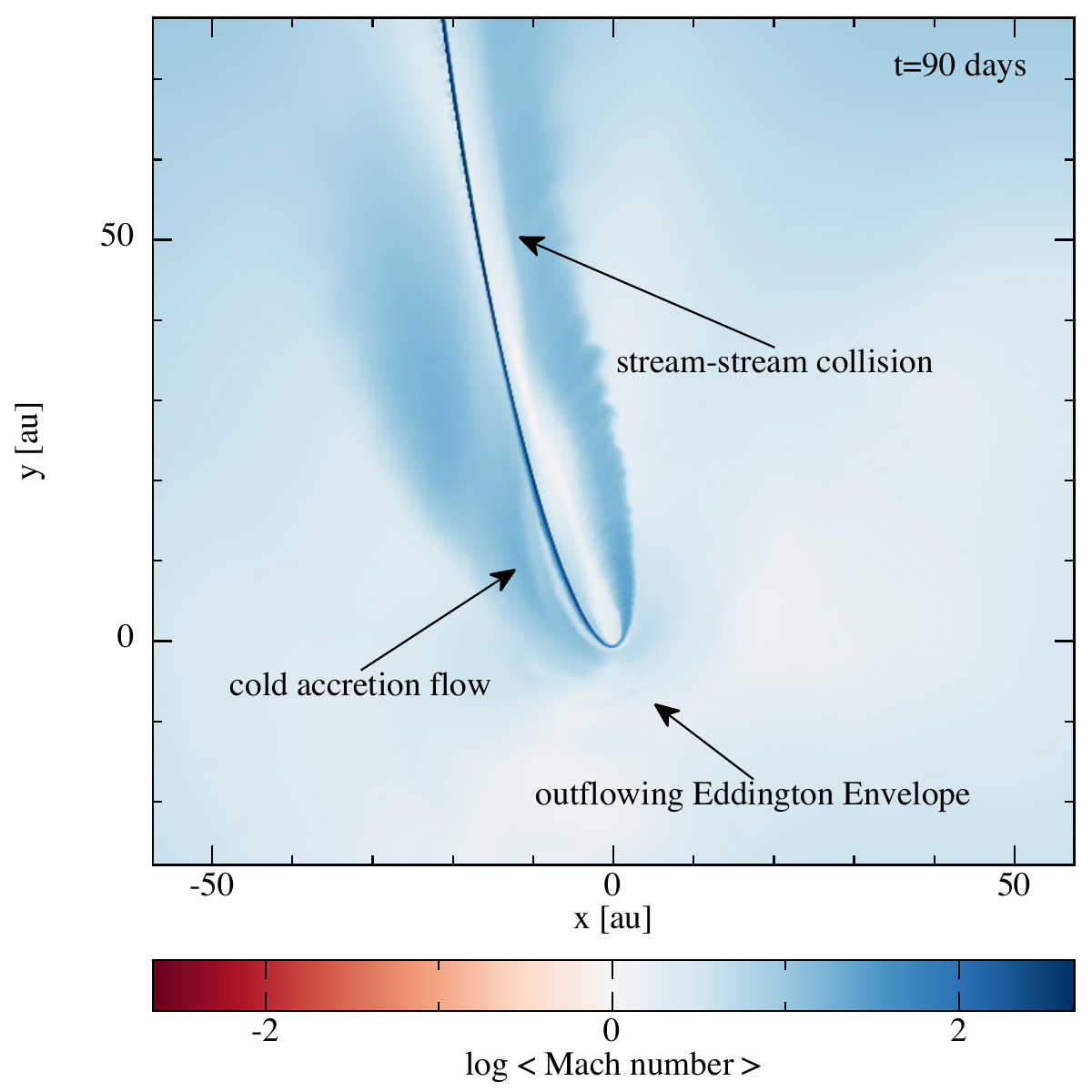}
\caption{Thermodynamics of the accretion flow, showing line-of-sight averaged internal energy $<u> \equiv \int \rho u \,{\rm d}z/ \int \rho \,{\rm d}z$ (left) and Mach number (right) in the adiabatic calculation from Figure~\ref{fig:main} (with $\beta = 1$, $a=0$). The stream itself remains cold apart from mild heating caused by the nozzle shock at pericentre passage and some shock heating caused by stream-stream collisions near apocentre. The outflowing Eddington Envelope is powered by the accretion flow towards the innermost regions near the black hole. Data and scripts used to create the figure are available on Zenodo:\dataset[doi:10.5281/zenodo.11438154]{https://doi.org/10.5281/zenodo.11438154}}
   \label{fig:thermo}
\end{figure*}

Our adiabatic simulations show lateral expansion of the stream in the nozzle shock that occurs during second periapsis passage compared to isentropic calculations. Material in the returning stream, with a narrow range of energies and angular momenta, is spread out along a range of trajectories because of heating through the nozzle shock. Some of the nozzle shock heating is overestimated, especially at low numerical resolution (see Figure~\ref{fig:resolution_study}). However, provided the Mach number remains high then material continues to follow the precessing geodesic orbits similar to those shown in Figure~\ref{fig:disc_formation}. Figure~\ref{fig:thermo} shows the line-of-sight averaged internal energy in our highest resolution $\beta = 1$ calculation (as shown in Figure~\ref{fig:main}). As the right panel of Figure~\ref{fig:thermo} shows, the Mach number drops to around 50 after the nozzle shock in our highest resolution calculation, which is sufficient to continue mostly along a geodesic trajectory.

 More important to the circularisation process is the `outer shock' due to self-intersection of the stream near apoapsis, labelled as `stream-stream collision' in the figure. Here, the Mach number drops to approximately unity in the shocked material (see right panel of Figure~\ref{fig:thermo}). Such material is then sprayed on a broad range of trajectories (as per Figure~\ref{fig:machfan}), in particular providing the `cold accretion flow' (see labels in Figure~\ref{fig:thermo}) towards the black hole, enabling prompt accretion, similar to the process envisioned by \citet{bonnerotlu19}.

The process of stream-stream collisions initiating the accretion flow towards the black hole is similar to that which occurs in isentropic calculations (Appendix~\ref{sec:disc}). The main difference in the adiabatic simulations is that material reaching the inner regions close to the black hole no longer remains cold, with the additional heating producing kinetic outflows (Figure~\ref{fig:thermo}).

\section{Photospheric evolution}
\label{sec:lightcurves}

\begin{figure*}
   \centering
      \includegraphics[width=\textwidth]{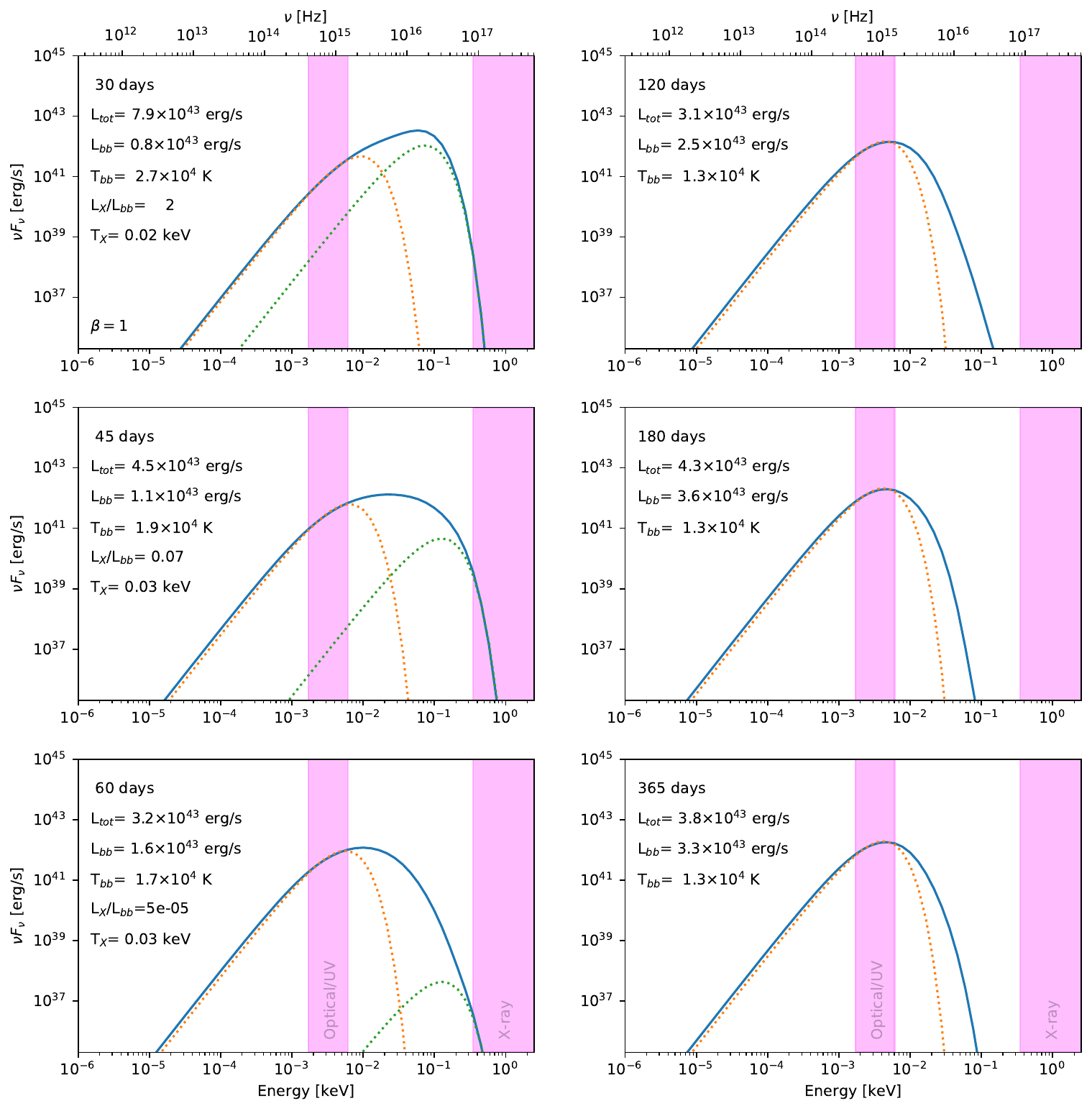}

   \caption{Synthetic spectral energy distributions computed from the $\beta = 1$ calculation with the observer in the $+z$ direction (blue curves). Orange dotted curves show single-temperature blackbody fits to the optical band, while green dotted curves show blackbody fits to the soft X-ray band. The legend lists the luminosities ($L_{\rm tot}$, $L_{\rm bb}$ and $L_{\rm x}$), the optical blackbody temperature ($T_{\rm bb}$) and the X-ray temperature.  Data and scripts used to create the figure are available on Zenodo:\dataset[doi:10.5281/zenodo.11438154]{https://doi.org/10.5281/zenodo.11438154}}
   \label{fig:spectra_beta1}
\end{figure*}

Figure~\ref{fig:spectra_beta1} shows our synthetic spectral energy distributions (SEDs) at selected times (see legend) for the $\beta = 1$ calculation. The overall spectrum is non-thermal, with significant excess at high energies compared to the optical blackbody fit (orange dotted curve; computed by fitting a single temperature blackbody to the optical band using \verb+scipy.optimize.curve_fit+). The high energy excess can be seen to produce soft X-rays whenever emission from the inner regions is visible. In the $\beta = 1$ calculation this occurs for $t \lesssim 61$ days, before the Eddington envelope smothers the central engine (see corresponding panels in Figure~\ref{fig:photosphere}).

While our predicted ratios of optical to X-ray emission of 1 to $10^{3}$ are similar to those observed, we find our predicted X-ray emission depends on $\beta$. Observations also show slightly higher X-ray temperatures in the range $0.05-0.15$ keV \citep{van-velzenetal21} and tend to find X-ray luminosities that increase with time.
Figure~\ref{fig:spectra_beta5} shows that something close to this occurs for the $\beta=5$ calculation. Here the more extreme precession leads to rapid smothering of the central engine in $t \lesssim 30$ days, but the faster stream dynamics means that the photosphere undergoes a visible retreat at $t \gtrsim 240$ days (see Figure~\ref{fig:lightcurves}) leading to increasing X-ray emission for $t \gtrsim 240$ days in the case where the star penetrates deeper.

\section{Effect of penetration factor}
\label{sec:beta5}

\begin{figure*}
   \centering
   \includegraphics[width=\textwidth]{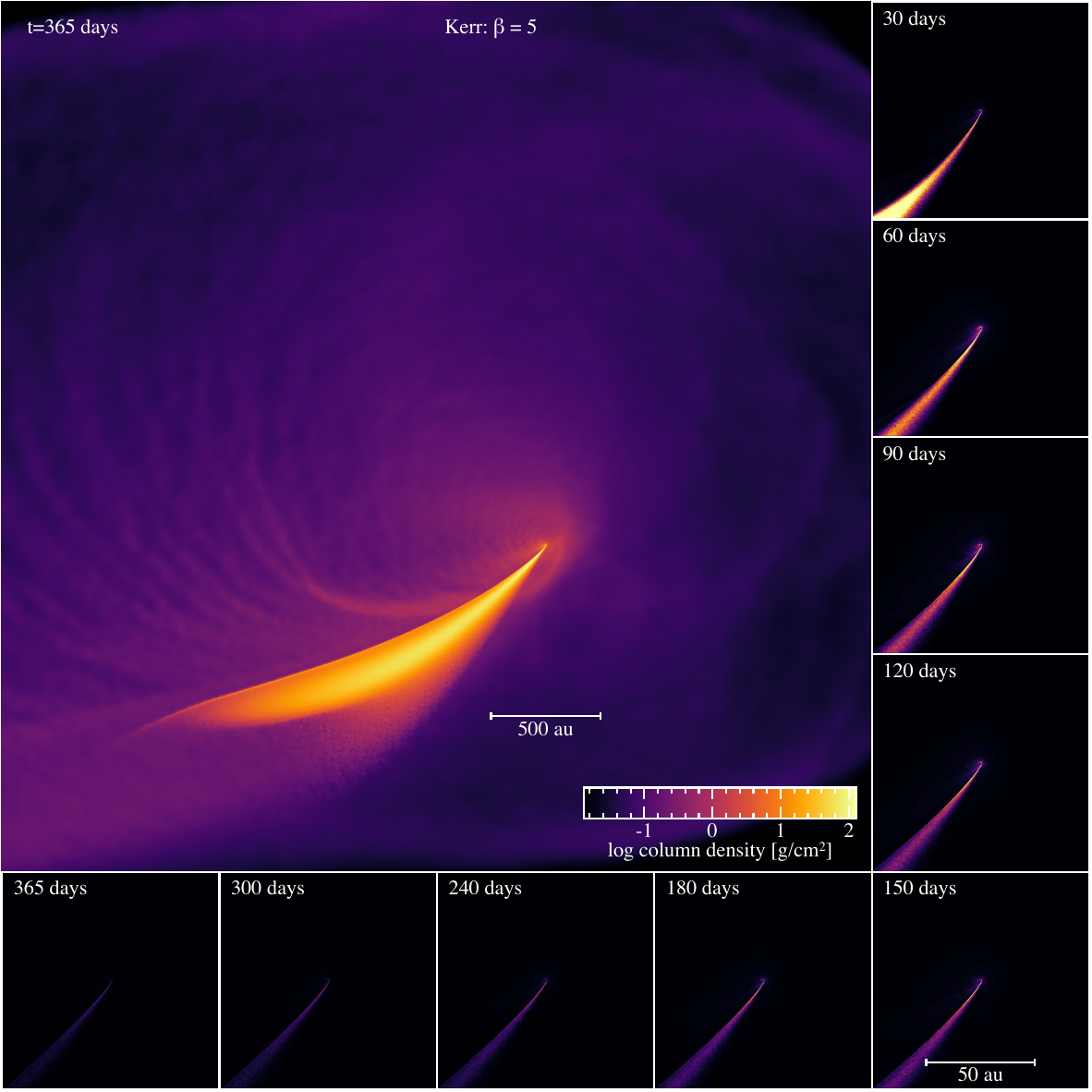}
   \caption{As in Figure~\ref{fig:main}, but for a star plunging closer to the black hole ($\beta = 5$). We show column density evolution for a $1M_\odot$ star on a parabolic orbit with $\beta=5$, disrupted by a $10^6M_\odot$ Kerr black hole with spin $a=0.99$. The initial orbital plane is misaligned by $60^\circ$ to the black hole spin plane. Main panel shows the debris after 365 days projected in the $x$-$y$ plane with log colour bar. Inset panels show stream evolution with a linear colour bar between 0 and 1500 g/cm$^2$. Data and scripts used to create the figure are available on Zenodo:\dataset[doi:10.5281/zenodo.11438154]{https://doi.org/10.5281/zenodo.11438154}}
   \label{fig:beta5}
\end{figure*}

Figure~\ref{fig:beta5} shows snapshots of the debris fallback for a calculation with an increased penetration factor of $\beta=5$, for a Kerr black hole with $a=0.99$ with the initial orbit inclined by 60$^\circ$ with respect to the black hole spin plane. 

Since the star approaches much closer to the black hole, the effects of apsidal and Lense-Thirring precession throw the star/debris stream into an orbital plane that is different to the initial one.
This changes the debris geometry projected in each coordinate plane, however the overall picture remains unchanged.
That is, the returning debris stream feeds the expansion of a low density Eddington envelope as in the $\beta=1$ case. Circularisation of the debris stream is faster in our $\beta = 5$ calculation because of the enhanced general relativistic precession, resulting in more prompt envelope formation (Figure~\ref{fig:beta5}).  

\begin{figure*}
   \centering
      \includegraphics[width=\textwidth]{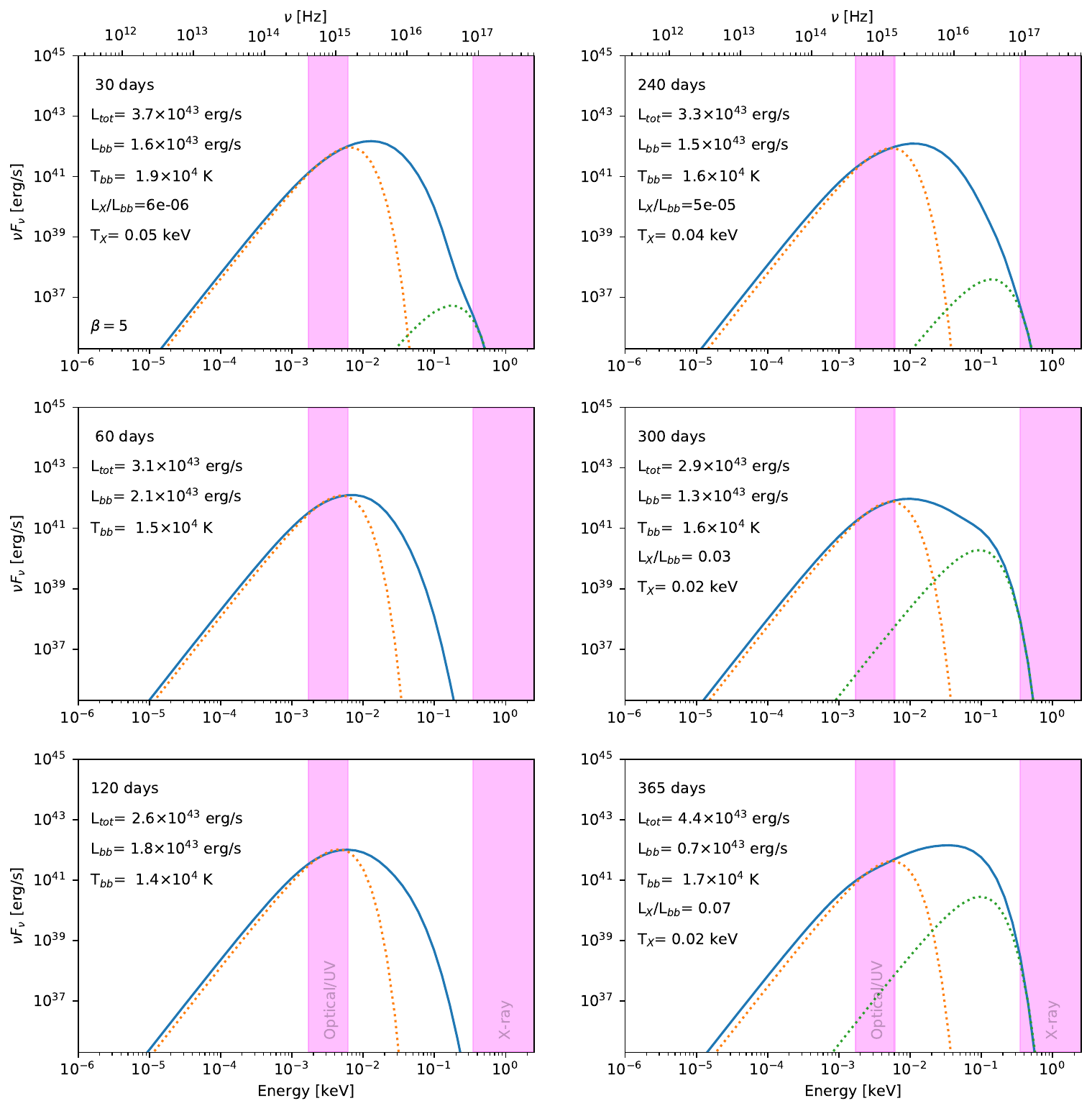}

   \caption{As in Figure~\ref{fig:spectra_beta1} but for the $\beta = 5$ calculation. Here the `bubble' is inflated on a shorter timescale, smothering the central engine in $t < 30$ days, but the retreat of the photosphere at $t \gtrsim 240$ days leads to soft X-ray emission as the hotter central regions become visible. Data and scripts used to create the figure are available on Zenodo:\dataset[doi:10.5281/zenodo.11438154]{https://doi.org/10.5281/zenodo.11438154}}
   \label{fig:spectra_beta5}
\end{figure*}

The right panel of Figure~\ref{fig:lightcurves} shows the corresponding lightcurves for this calculation (discussed in the main text), while Figure~\ref{fig:spectra_beta5} shows the corresponding spectral energy distributions.

\section{Effect of resolution}\label{sec:resolution}
\begin{figure*}
   \centering
   \includegraphics[width=\textwidth]{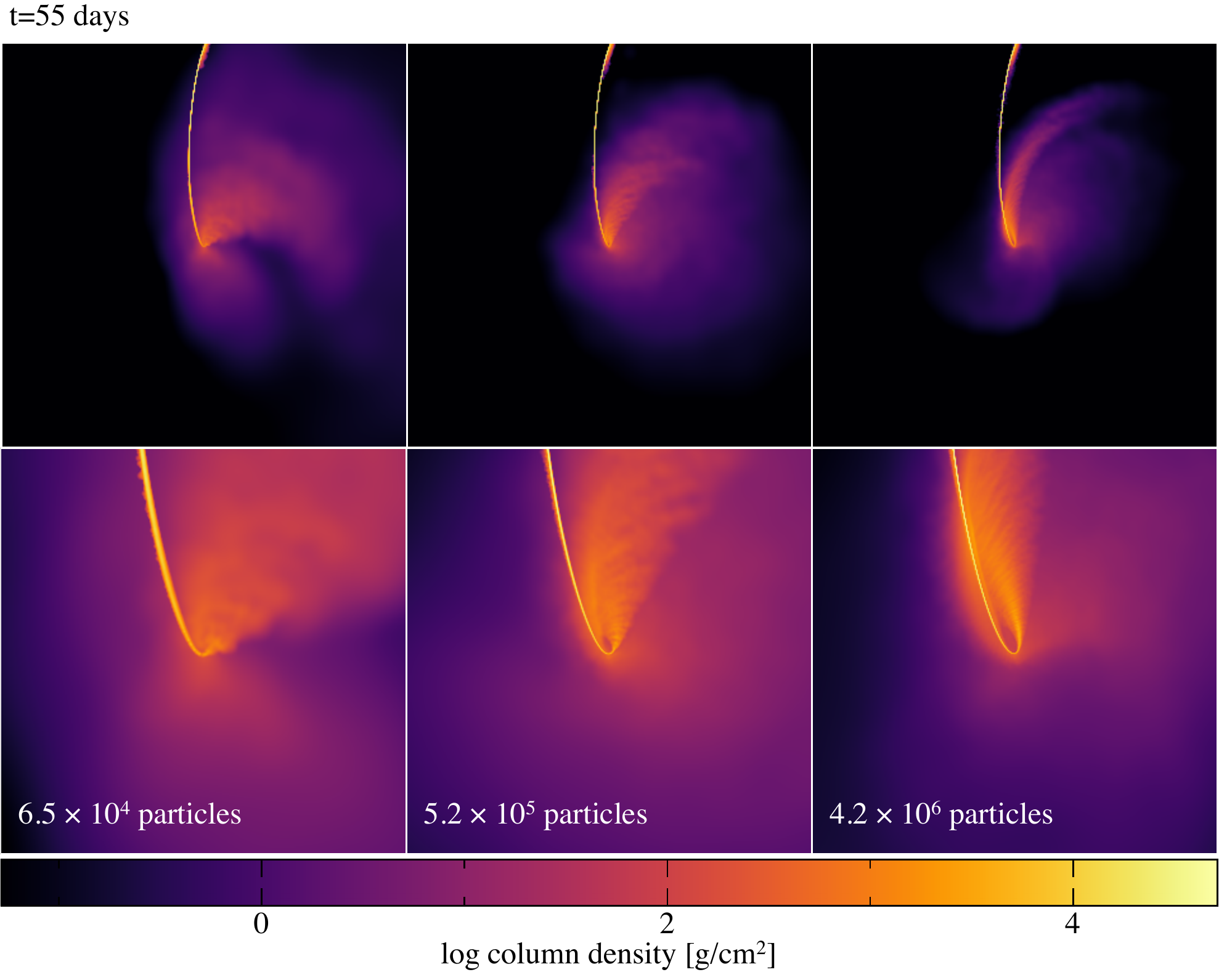}
   \caption{Effect of resolution in calculations with $a=\theta=0$ and $\beta=1$, as in Figure~\ref{fig:main}. \textit{Top row:} comparison of the large scale structure, where each panel is 500 au $\times$ 500 au. \textit{Bottom row:} comparison of the stream fanning at small scales where each panel is 100 au $\times$ 100 au. The total number of particles increases by approximately a factor of 8 between columns from left to right, with the third column corresponding to the resolution used in our main calculations. Data and scripts used to create the figure are available on Zenodo:\dataset[doi:10.5281/zenodo.11438154]{https://doi.org/10.5281/zenodo.11438154}}
   \label{fig:resolution_study}
\end{figure*}

Figure~\ref{fig:resolution_study} shows the effect of resolution in our simulations. The setup is identical to that shown in Figure~\ref{fig:main}.
The \textit{top} row compares the large scale structure, while the \textit{bottom} compares the details of the stream on a smaller scale as it passes through the nozzle shock.
We increased the number of particles by approximately a factor of 8 between each column from left to right, corresponding to a factor of 2 increase in spatial resolution.
 We find that heating through the nozzle shock is overestimated at low numerical resolutions, leading to a more isotropic outflow. While the degree of nozzle shock heating is not fully converged even at our highest resolution (as expected from \citealt{bonnerotlu22}), the Mach number remains sufficiently high that the material mostly follows colder trajectories along orbital geodesics as in the isentropic case (see Figure~\ref{fig:disc_formation}). Hence the process of circularisation tends towards being driven mainly by the  stream self-intersection at apocentre rather than nozzle shock heating at pericentre which is dominant at low resolution. Our results also imply that the envelope may be less isotropic in reality, with more material in a `spiral arm' that is denser than the rest of the envelope.

To clarify the numerical dependence of the lateral stream expansion in the nozzle shock we performed two additional numerical experiments. The first involved a simple injection of material at different Mach numbers into an otherwise empty domain with no central black hole. We chose parameters typical of those observed in the tidal disruption calculations, by injecting material in a stream at the typical location of the returning stream just prior to pericenter in our calculations, namely at $(x,y)=(105,-150)$ in geometric units. We fixed the injection velocity to be the incoming stream velocity observed in our simulations, namely at $v = 0.5 \sqrt{GM/R_p}$ where $R_{\rm p}$ is the pericenter distance for the original star, at $R_{\rm p} = R_{\rm t} = 47.1 GM/c^2$. We injected a stream as a cylinder of particles with a radius of $10 R_{\odot}$ and varied the Mach number of the stream (by setting the incoming thermal energy $u$). Particles were placed in a cubic lattice arrangement, cropped to the radius of the cylinder. Calculations were performed with an adiabatic equation of state with $\gamma = 5/3$. We adopted a resolution of 16 particles per stream width, giving 565,940 particles in the domain at $t=100$. The mass of the SPH particles is arbitrary for this problem (we chose $m_{\rm part} = 10^{-12} M$) and we employed the non-relativistic version of {\sc Phantom} for simplicity.

\begin{figure*}
   \centering
   \includegraphics[width=\textwidth]{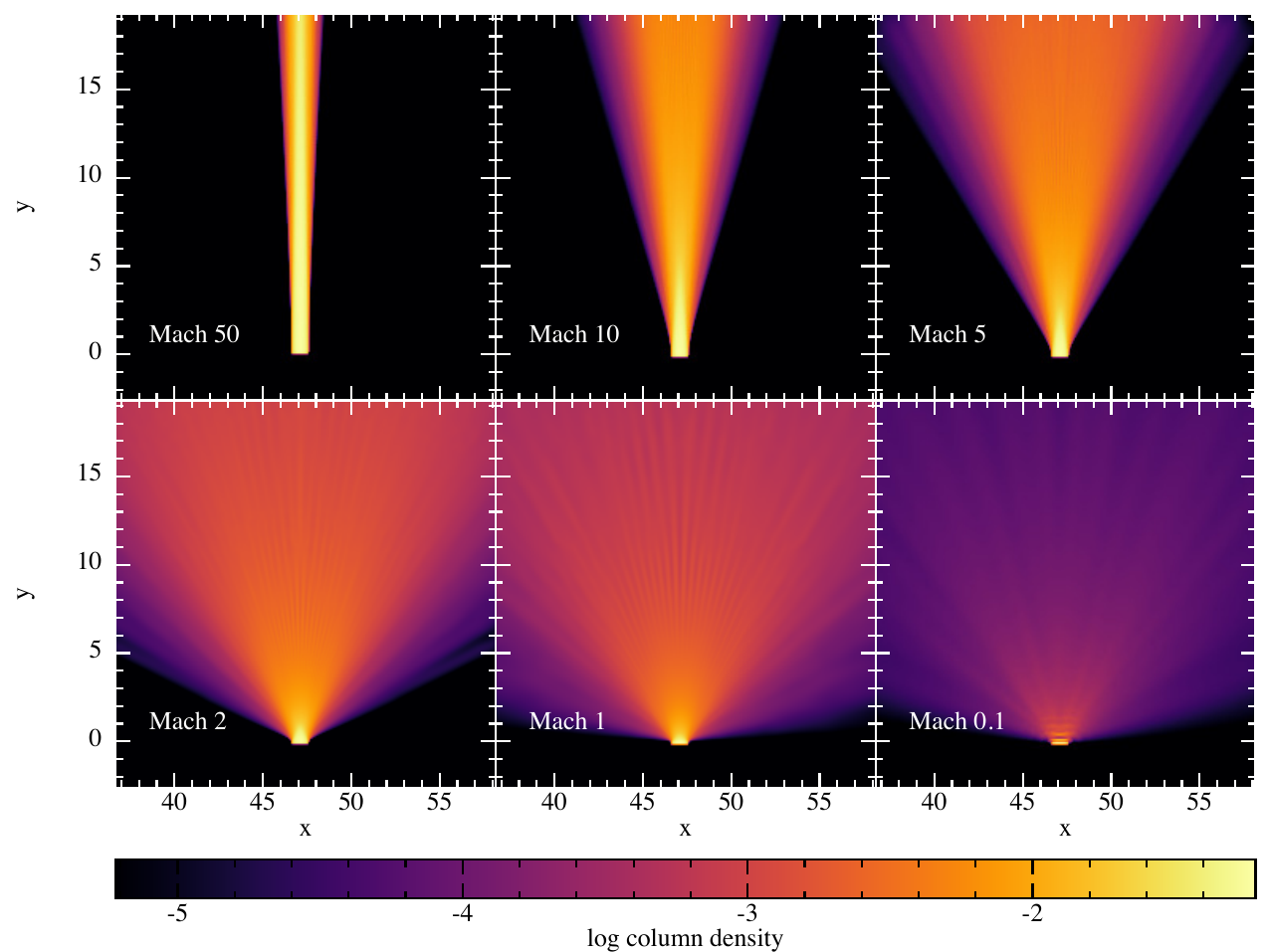}
   \caption{Numerical experiment where a stream is injected into an otherwise empty domain with a fixed injection speed but varying Mach numbers. Lateral expansion of the stream can be seen to depend solely on the Mach number of the flow. Data and scripts used to create the figure are available on Zenodo:\dataset[doi:10.5281/zenodo.11438154]{https://doi.org/10.5281/zenodo.11438154}}
   \label{fig:machfan}
\end{figure*}

Figure~\ref{fig:machfan} shows the results of this experiment. The lateral expansion of the stream can be seen to be a simple function of the Mach number of the flow. Below Mach 10 significant lateral expansion of the stream may be observed. We found the degree of lateral expansion to be independent of both the injected stream width and the numerical resolution, depending solely on the Mach number.
 
Our second experiment was the same but with a point mass placed at the origin with $M = M_{\rm BH} = 10^6 M_{\odot}$ to capture the pericenter dynamics. Since we injected material with a finite width but a single velocity the infinite Mach number orbital trajectories of the highly eccentric orbits cause a `traffic jam' at pericenter which induces P{\rm d}V work on the stream. We adopted an initial Mach number of 50, consistent with that measured from our tidal disruption calculations, but set the initial radius of the injected stream to $R = 1, 2$ and $10 R_{\odot}$, respectively. Apart from the overall drop in density as the stream width is increased, the lateral expansion of the stream remained the same in all cases. This demonstrates that an artificially wide or unresolved stream width is not the main cause of the lateral expansion.

\begin{figure*}
   \centering
   \includegraphics[width=\textwidth]{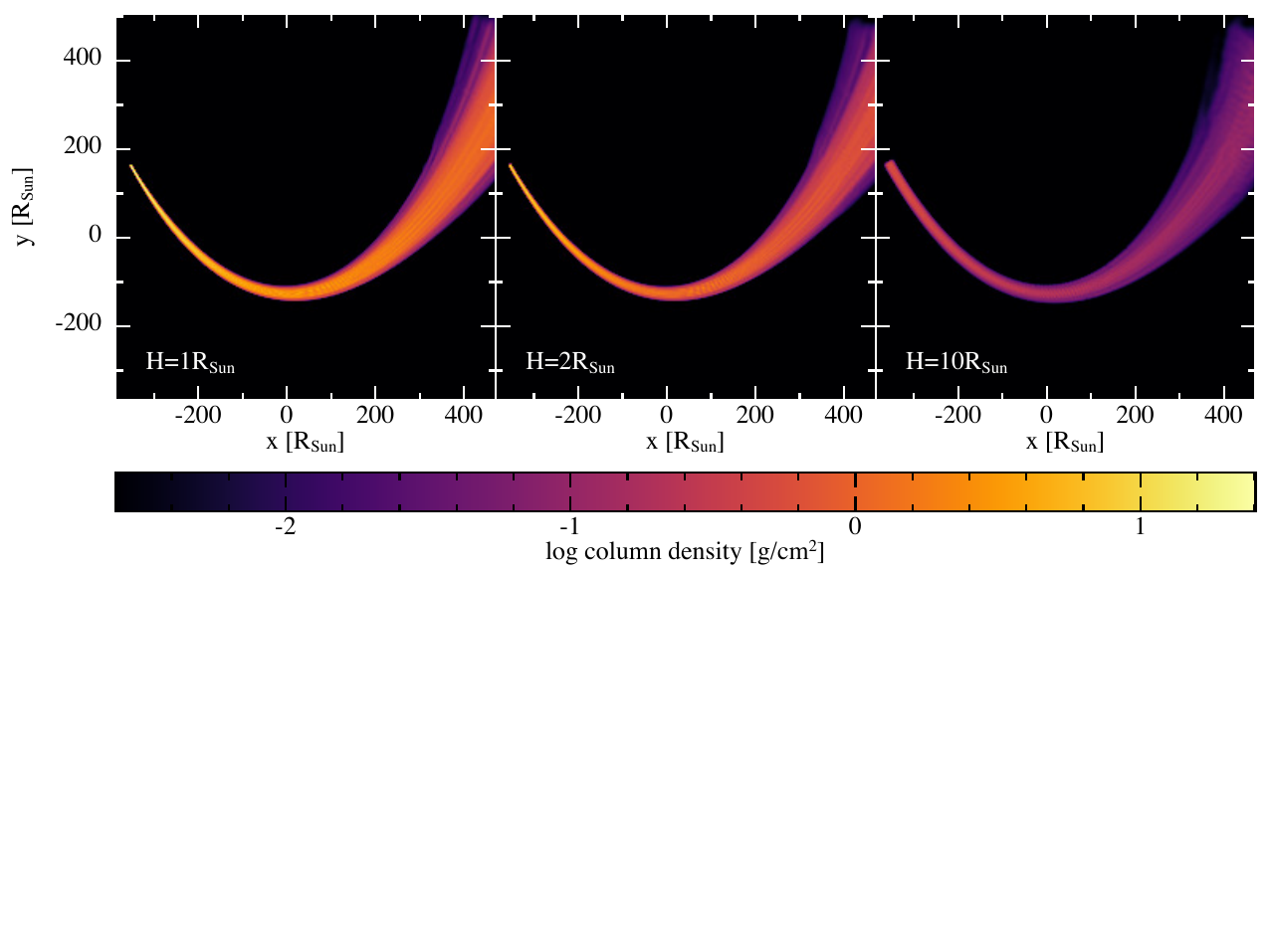}
   \vspace{-5cm}
   \caption{As in Figure~\ref{fig:machfan} but with a point mass placed at the origin, and experimenting with different initial stream radii $H$ for the same initial Mach number of $M=50$. Data and scripts used to create the figure are available on Zenodo:\dataset[doi:10.5281/zenodo.11438154]{https://doi.org/10.5281/zenodo.11438154}}
   \label{fig:orbitfan}
\end{figure*}        

Figure~\ref{fig:resolution_study} shows the post-pericentre Mach number drop is overestimated at low resolution due to excess heating in the nozzle shock, leading to an overestimated stream width. The qualitative behaviour is unchanged so long as the post-pericentre Mach number remains high, as occurs in our highest resolution adiabatic calculations, and isentropic calculations at all resolutions.

Conservation properties are well preserved by the numerical scheme at all resolutions. For example, in the calculation shown in Figure~\ref{fig:main} the energy error is $\Delta E/E \sim 6 \times 10^{-4}$ up to $\sim 20$ days when accretion commences, after which the total energy rises steadily due to the removal of accreted mass with negative specific energy from the simulation. Total angular momentum is conserved to $\Delta L/L \sim 2\times 10^{-4}$ (0.02\%) over the entire 365 days of the simulation.

\end{document}